\documentclass[aps,prb,amsmath,superscriptaddress,showpacs,longbibliography]{revtex4-1}
\usepackage{color}
\usepackage{amssymb}
\usepackage{dcolumn}
\usepackage{amstext}
\usepackage{graphicx}
\usepackage{amsmath}
\usepackage{tabularx}
\usepackage{amsthm}
\usepackage{multirow}
\usepackage{longtable}
\usepackage{amsfonts}
\usepackage{amssymb}
\usepackage{latexsym}
\usepackage{bbold}
\usepackage{float}
\usepackage[colorlinks=true,urlcolor=blue,anchorcolor=blue,linkcolor=red,citecolor=red,breaklinks=true]{hyperref}
\usepackage{adjustbox}
\usepackage{ulem}
\def\k{{\bf k}}
\def\rw{\tilde{\omega}}

\def\rQ{\tilde{Q}}
\def\dv{{\bf d}}

\def\ute2{UTe$_2$}
\def\d2h{$D_{2h}$}

\def\q{{\bf q}}
\def\bsig{\boldsymbol{\sigma}}
\def\bSig{\boldsymbol{\Sigma}}
\definecolor{applegreen}{rgb}{0.55, 0.71, 0.0}
\definecolor{magneta}{rgb}{1.0,0.0,1.0}

\begin{document}

\title{Thermal conductivity of nonunitary triplet superconductors: application to UTe$_2$}

\author{Vivek Mishra}
\affiliation{Department of Physics, University of Florida, Gainesville, Florida 32611 }

\author{Ge Wang}
\affiliation{Department of Physics, University of Florida, Gainesville, Florida 32611 }

\author{P.  J. Hirschfeld}
\affiliation{Department of Physics, University of Florida, Gainesville, Florida 32611 }

\date{\today}

\begin{abstract}
There is considerable evidence that the heavy fermion material UTe$_2$ is a spin-triplet superconductor, possibly manifesting time-reversal symmetry  breaking, as measured by Kerr effect  below the critical temperature, in some samples.  Such signals can arise due to a chiral orbital state, or possible nonunitary pairing.  Although experiments at low temperatures appear to be consistent with point nodes in the spectral gap, the detailed form of the order parameter and even the nodal positions are not yet determined.  Thermal conductivity measurements can extend to quite low temperatures, and by  varying the heat current direction should be able to provide information on the order parameter structure.   Here we derive a general expression for the thermal conductivity of a spin triplet superconductor, and use it to compare the low-temperature behavior of various states proposed for UTe$_2$.
\end{abstract}

\date{\today}

\maketitle

\section{Introduction}\label{sec:intro} 
The uranium-based superconductor $\mathrm{UTe}_2$ has stimulated a large number of experimental and theoretical studies, initially because of its apparent role as a paramagnetic end-point of a family of ferromagnetic superconductors\citep{Ran2019,Aoki2019,Aoki_2022_Review}, and later as evidence for spin-triplet superconductivity accumulated. {The nuclear magnetic resonance (NMR) Knight shift measurements on the earlier samples did not show any change in the superconducting state\cite{Fujibayashi2022}, although recent Knight shift measurements on high quality samples show a small reduction along all three axes\cite{Matsumura2023}. Both measurements support spin triplet pairing, however, the spin structure of Cooper pairs remains unclear. Another piece of evidence that indicate spin triplet pairing is the size of the upper critical field $H_{c 2}$ that exceeds the Pauli limit for all field directions\cite{Ran2019,Aoki2019a}.} Measured power law temperature dependence in NMR  relaxation, specific heat\citep{Ran2019} and thermal conductivity\citep{Metz2019,Hayes2024}   were found to be consistent with point nodes, expected for a triplet superconductor in a system with strong spin-orbit coupling\citep{VolovikGorkov1984,UedaRice1985,Blount1985}.  Finally,  a reentrant superconducting phase was shown to be stabilized at  high magnetic fields\citep{Ran2019B}.
    
    A second set of measurements relevant to the nature of the superconducting state purports to exhibit  evidence for time reversal symmetry breaking (TRSB) below $T_c$, suggesting that UTe$_2$ may support the long-sought  chiral $p$-wave state that may serve as a quantum computing platform\citep{Madhavan20,Kitaev2001,Wilczek2009, Ivanov2001,SCZ2011}.  Initially,  polar Kerr effect\citep{Hayes2021} experiments suggested TRSB occurred in the superconductor, implying the existence of a multicomponent spin triplet order parameter.  According to group theoretical classifications of the 1-dimensional irreducible representations (irreps) allowed in orthorhombic symmetry, order parameters corresponding to single irreps must be unitary triplet states, meaning that any TRSB must arise from a nonunitary multicomponent state.  Such combinations of 1D representations were discussed intensively, particularly because a double specific heat transition was sometimes seen in early samples, recalling the specific heat experiments in multicomponent UPt$_3$.

More recently, measurements of a new generation of high quality UTe$_2$ crystals grown in molten salt flux  have challenged this characterization of UTe$_2$ as a chiral triplet state breaking time reversal symmetry.  The Kerr effect was observed in a sample showing two specific heat jumps, but as the quality of the samples improved, only a single transition was observed\citep{Rosa2022,Sakai2022}.   A recent Kerr effect investigation of both the old and new generation $\mathrm{UTe}_2$ single crystals displaying a single  specific heat jump found no evidence for TRSB superconductivity\citep{Ajeesh2023}. Similarly, muon spin relaxation ($\mu$SR) measurements of the molten-salt flux grown samples found no evidence of TRSB\citep{Azari2023}.  Finally,  sound velocity changes  across $T_c$\citep{Theus2023}, as well as  recent NMR Knight shift measurements on similar samples\citep{Matsumura2023} both point to a single-component, odd parity  order parameter, i.e. are inconsistent with the previous hypothesis of nonunitary pairing.
    
The thermal conductivity $\kappa(T)$ is an important  probe of  the gap structure of unconventional superconductors, reflecting the ability of the superconductor to carry heat current in various directions. The theory of thermal conductivity  in {\it unitary} triplet superconductors is quite similar to the well-known theory developed  for singlet superconductors\citep{AmbegaokarGriffin:1994,KadanoffMartin1963}.  Most of the popular model triplet states in the literature, including the $^3$He-A phase, are in this class.  In that case, the triplet quasiparticle energies are $E_{\bf k}=\sqrt{\xi_{\bf k}^2+|{\bf d}({\bf k})|^2}$, where ${\bf d}({\bf k})$ is the triplet order parameter vector defining its structure in spin space via $\Delta_{\sigma\sigma'} =  \left[  {\bf d}({\bf k})\cdot \vec\sigma (i\sigma_2)\right]_{\sigma\sigma'}$.  
    Since the thermal current response depends only on the quasiparticle energies, and the same expressions can be used for triplet superconductors\citep{SSchmitt-Rink:1986,PJHirschfeld:1986,PJHirschfeld:1988} with $|\Delta({\bf k})|^2$ replaced by $|{\bf d}({\bf k})|^2$.    As we show below, however, in the  nonunitary case, additional terms involving the spin moment  ${\bf q}\equiv i {\bf d}({\bf k}) \times {\bf d}^*({\bf k})$ carried by quasiparticles of momentum $\bf k$ occur in both the quasiparticle energies and the weights of scattering processes.  Furthermore, in  nonunitary triplet superconductors, the zeros of $|{\bf d}({\bf k})|^2$ differ from those of the zeros of $E_{\bf k}$ even when  ${\bf k}$ is on the  Fermi surface  (``spectral nodes").  This distinction may be important;  it was suggested  by  Ishihara et al.\citep{Ishihara2023} that in UTe$_2$, complex linear combinations of 1D irreducible representations could  support spectral nodes pointing in generic directions in the orthorhombic Brillouin zone and thereby explain early experiments exhibiting TRSB and relative isotropy of the low-temperature penetration depth.   On the other hand,  order parameters   corresponding to a single 1D irrep must be real, with nodes along high symmetry axes.  

In this paper, we derive a general form of the thermal conductivity of a triplet superconductor in the presence of nonmagnetic disorder, and evaluate it for various types of triplet states that have been proposed for UTe$_2$. The aim is to see if there are qualitative distinctions between the thermal conductivity temperature and heat current direction dependence of unitary and nonunitary states, and whether or not predictions of low-temperature behavior can be used, by comparison with experiments, to identify the ground state of UTe$_2$.

\section{Model \& Formalism}\label{sec:mod_form}
\begin{table}[tbph]
    \centering
    \begin{tabular}{ccc}
    \hline
    \hline
         $\Gamma$ & Gap function $\dv(\k)$   & Nodes  \\
         \hline
         $A_{1u}$ &  ($p_1 k_x, p_2 k_y, p_3 k_z$)   & accidental \\  
         $B_{1u}$ &  ($p_1 k_y, p_2 k_x, p_3 k_x k_y k_z $)   &  $z$ axis \\
         $B_{2u}$ & ($p_1 k_z, p_2 k_x k_y k_z, p_3 k_x $)   & $y$ axis \\
         $B_{3u}$ & ($p_1 k_x k_y k_z,  p_2 k_z, p_3 k_y $)  & $x$ axis\\
    \hline
    \hline
    \end{tabular}
    \caption{List of possible spin-triplet superconducting states for an orthorhombic crystal with strong spin orbit coupling. Here $p_{i=1,..,3}$ are constants and $\forall  p_i \in \mathbf{R}$.}
    \label{Table1}
\end{table}
\begin{table*}
\centering
\begin{tabular}{|c|c|c|c|c|c|}
\hline
Classification &  $\dv(\k)$    & \multicolumn{2}{|c|}{Cylindrical FS} & \multicolumn{2}{|c|}{Spherical FS}   \\ \cline{3-6}
& & $r$ & N$_{nodes}$ &  $r$ & N$_{nodes}$ \\
\hline
\multirow{16}{*}{\shortstack{Antiferromagnetic\\ non-unitary}} & \multirow{4}{*}{$A_{1u} + i r B_{1u}$} & $r <1 $  & 0   & $r<1$  & 0   \\ 
&& $r=1$  & 4  ($xy$ plane)   &  $r=1$  & 4  ($xy$ plane) \\ 
&& $1<r\leq \sqrt{2}$  & 8  ($xz$, $yz$ planes)   &  $r>1$  &   8  ($xz$, $yz$ planes)  \\ 
&& $r> \sqrt{2}$  & 0  &   &  \\ \cline{2-6} 
& \multirow{4}{*}{$A_{1u} + i r B_{2u}$} & $r < 1$  & 0   & $r<1$  & 0   \\ 
&& $r=1$  & 2  ($x$ axis)   &  $r=1$  & 4  ($xz$ plane) \\ 
&& $1<r\leq \sqrt{2}$  & 4  ($xy$ plane)   &  $r>1$  &   8  ($xy$, $yz$ planes)  \\ 
&& $r > \sqrt{2}$  & 8  ($xy$, $yz$ planes)   &  &    \\ \cline{2-6} 
& \multirow{4}{*}{$A_{1u} + i r B_{3u}$} & $r < 1$  & 0   & $r<1$  & 0   \\ 
&& $r=1$  & 2  ($y$ axis)   &  $r=1$  & 4  ($yz$ plane) \\ 
&& $1<r\leq \sqrt{2}$  & 4  ($xy$ plane)   &  $r>1$  &   8  ($xy$, $xz$ planes)  \\ 
&& $r > \sqrt{2}$  & 8  ($xy$, $xz$ planes)   &  &    \\ \cline{1-6} 
\multirow{15}{*}{\shortstack{Ferromagnetic \\ non-unitary}} & \multirow{4}{*}{\shortstack{$B_{1u} + i r B_{2u}$ \\ ( $\hat{x}$ spin moment )}} & $r <\frac{1}{\sqrt{2}} $  & 0   & $r<1$  & 4 ($xz$ plane)   \\
&& $\frac{1}{\sqrt{2}}\leq r <1$ & 4 ($xz$ plane) & $r=1$ & 2 ($x$ axis) \\
&& $r=1$ & 2 $x$ axis & $r>1$ &  4 ($xy$ plane) \\
&& $r>1$ & 4 ($xy$ plane) & &  \\ \cline{2-6}
& \multirow{4}{*}{ \shortstack{$B_{1u} + i r B_{3u}$ \\ ( $\hat{y}$ spin moment )}} & $r <\frac{1}{\sqrt{2}} $  & 0   & $r<1$  & 4 ($yz$ plane)   \\
&& $\frac{1}{\sqrt{2}}\leq r <1$ & 4 ($yz$ plane) & $r=1$ & 2 ($y$ axis) \\
&& $r=1$ & 2 ($y$ axis) & $r>1$ &  4 ($xy$ plane) \\
&& $r>1$ & 4 ($xy$ plane) & &  \\ \cline{2-6}
& \multirow{3}{*}{\shortstack{$B_{2u} + i r B_{3u}$ \\ ($\hat{z}$ spin moment )}} & $r \leq \frac{1}{\sqrt{2}} $  & 4  ($yz$ plane)  & $r<1$  & 4 ($yz$ plane)   \\
&& $\frac{1}{\sqrt{2}}< r <\sqrt{2}$ & 0 & $r=1$ & 2 ($z$ axis) \\
&& $r\geq \sqrt{2}$ & 4 ($xz$ plane) & $r>1$ & 4 ($xz$ plane)  \\
\hline
\end{tabular}
\caption{Six possible mixed IR states}
\label{Table2}
\end{table*}
\subsection{Superconducting State}\label{subsec:SC}
In a general triplet superconductor, the spin-structure of the superconducting order parameter is constrained by the underlying crystal symmetries. The structure of the  \ute2 crystals corresponds to  the orthorhombic point group \d2h, and the symmetry of the odd-parity pairing states can be deduced  according to the irreducible representations of the \d2h point group\citep{Marchenko1986,Sigrist1987,Annett1990}. Table \ref{Table1} shows the odd parity triplet superconducting states that we consider in this article. Here, we do not consider the even parity states for the \d2h point group, and we further assume strong spin-orbit coupling (SOC) due to heavy atoms like U and Te. In the weak SOC limit, the odd-parity states for the \d2h point group come with line nodes that are not consistent with experimental measurements. 

The $\dv$-vector is real for the superconducting states listed in Table \ref{Table1}, hence these states preserve time reversal symmetry (TRS). These states are unitary triplet states, \textit{i.e.} $\hat{\Delta}^\dagger \hat{\Delta}\propto \hat{\mathbb{1}}$. We denote a 2$\times$2 matrix in the spin space with $\hat{\square}$ and a 4$\times$4 matrix in the Nambu-spin space with $\check{\square}$. A TRSB state is not possible with a single component order parameter, noting that  the \d2h group has only one dimensional irreducible representations. We construct the TRSB superconducting state with a combination of two irreducible representations, and all such possible combinations are shown in Table \ref{Table2}. In principle, a combination of two different irreducible representations involves six real constants, however, we introduce a single parameter model for the TRSB states. The effective $\dv$-vector is $(\dv_1 +i r \dv_2)/\sqrt{1+r^2}$, where $r$ is the mixing parameter that determines the relative strength of the individual order parameter. The individual $\dv$-vectors are listed in Table \ref{Table1} with all the coefficients $p_{i=1,2,3}$ are set to unity.

In the Nambu-spin basis, the mean field Hamiltonian reads,
\begin{equation}
\check{\mathbf{H}} =\begin{pmatrix}
\xi_\k \sigma_0 & i \Delta_0(\mathbf{d} \cdot \boldsymbol{\sigma}) \sigma_y \\
-i \Delta_0 \sigma_y\left(\mathbf{d}^* \cdot \boldsymbol{\sigma}\right) & -\xi_\k \sigma_0
\end{pmatrix},
\label{Eq:H_mf}
\end{equation}
where $\xi_\k$ is the electronic dispersion, $\Delta_0$ is the superconducting gap energy scale, $\boldsymbol{\sigma}$ is the Pauli vector in the spin space spanned by the
Pauli matrices, and $\sigma_0$ is the identity matrix in the spin space. {We adopt a model where the } electronic dispersion reads,
\begin{eqnarray}
\xi_\k = \frac{\hbar^2 k_x^2}{2m_a}+  \frac{\hbar^2 k_y^2}{2m_b}-\mu -2 t_\perp \cos k_z,
\end{eqnarray}
where $m_{a/b}$ is the effective masses along $\hat{x}/\hat{y}$ directions, $t_\perp$ is the hopping energy that controls the $\hat{z}$ velocity, and $\mu$ is the chemical potential.  We  further assume $t_\perp \ll \mu$. An alternate dispersion with closed Fermi surface has been considered in the appendix.  The bare Green's function  is,
\begin{equation}
\check{G}_0=(\check{\mathbb{1}} \omega-\check{\mathbf{H}})^{-1}=\begin{pmatrix}
\hat{G}_{11} & \hat{G}_{12} \\
\hat{G}_{21} & \hat{G}_{22}
\end{pmatrix}.
\end{equation}
Here, $\omega$ is the quasi-particle energy. The Matsubara Green's function can be obtained by $\omega \rightarrow i \omega_n$. The 2$\times$2 matrices in the spin space are,
\begin{eqnarray}
\hat{G}_{11} &=& \frac{(\omega + \xi)}{{D}} \left[ (\omega^2-\xi^2-\Delta_0^2 |\dv|^2)\sigma_0 + \Delta_0^2 \mathbf{q}\cdot \boldsymbol{\sigma} \right], \\
\hat{G}_{12} &=& \left[ (\omega^2-\xi^2-\Delta_0^2 |\dv|^2)\sigma_0 + \Delta_0^2 \mathbf{q}\cdot \boldsymbol{\sigma} \right]  \frac{i\Delta_0 (\dv\cdot \boldsymbol{\sigma}) \sigma_y }{{D}} , \\
\hat{G}_{21} &=&  -\left[ (\omega^2-\xi^2-\Delta_0^2 |\dv|^2)\sigma_0 + \Delta_0^2 \mathbf{q}\cdot \boldsymbol{\sigma}^T \right] \frac{\Delta_0 i\sigma_y (\dv^* \cdot \boldsymbol{\sigma})}{{D}}   , \\
\hat{G}_{22} &=& \frac{(\omega - \xi)}{{D}} \left[ (\omega^2-\xi^2-\Delta_0^2 |\dv|^2)\sigma_0 + \Delta_0^2 \mathbf{q}\cdot \boldsymbol{\sigma}^T \right].
\label{Eq:G0}
\end{eqnarray}
Here $\q=i (\dv \times \dv^*)$ and  the denominator ${D}$ is,
\begin{eqnarray}
{D} &=& \left( \omega^2 -\xi^2 - \Delta_0^2 |\dv|^2\right)^2 - \Delta_0^4 |\mathbf{q}|^2 = (\xi^2-\omega^2+\Delta^2_+) (\xi^2-\omega^2+\Delta^2_-).
\label{Eq:Dnom}
\end{eqnarray}
Here we introduce $\Delta^2_\pm = \Delta_0^2 (|\dv|^2\pm|\q|)$. For the unitary case, $\q=0$, therefore, there is only a single energy scale. In contrast, for the TRSB non-unitary states, $\q\neq 0$, which leads to non-degenerate excitation energies. 
For single component order parameters, $\q$ vanishes. However, for a mixture of multiple irreducible representations $\q$ remains finite, and it can be interpreted as spin moment of the Cooper pairs. The average of $\q$ over the Fermi surface may or may not vanish. The non-unitary states can therefore be further divided in anti-ferromagnetic (AF) and ferromagnetic (FM) states, where the average of $\q$ vanishes over the Fermi surface for the former, while remains finite for the latter\citep{VolovikGorkov1984}. The six possible  nonunitary states are shown in Table \ref{Table2}, with the possibility of nodes on a spherical {or} a cylindrical Fermi surface open along the $\hat{z}$ axis. For cylindrical Fermi surface, we adopt cylindrical coordinates with $k_z$ dependence of the  gap functions replaced with $\sin (k_z d /2 )$, where $d$ is the $z$-axis. The factor of half is added to ensure only a single pair of point nodes in the first Brillouin zone in the unitary limit. However, this does not have any qualitative effect on our results. 

\subsection{Impurity Scattering \& Thermal Transport}\label{subsec:kappa_b}
In order to calculate the thermal conductivity, we need to include the effect of impurity scattering that dominates all other relaxation mechanism at low temperatures. We consider elastic impurity scattering due to pointlike defects, and include its effect through a disorder-averaged self-energy. The impurity self-energy is calculated within the self-consistent T-matrix approximation. The momentum integrated Green's function ${\hat g}$ has both vector and scalar components for the normal Green's functions,
\begin{eqnarray}
\hat{g}_{11}&=& \pi N_0 \left(g_0  + \mathbf{g}\cdot \bsig \right),  \label{Eq:g0a} \\
\hat{g}_{22}&=& \pi N_0 \left(g_0 + \mathbf{g}\cdot \bsig^T \right),
\label{Eq:g0b}
\end{eqnarray}
and the anomalous Green's functions $\hat{g}_{12}$ and $\hat{g}_{21}$  vanish because the odd-parity order parameter averages to zero. In Eq. \eqref{Eq:g0a} and \eqref{Eq:g0b},  $\mathbf{g}$ is directly related to the Fermi surface average of the spin moment $\q$ and remains finite for the chiral non-unitary states, only. Using these integrated Green's function, we can write the T-matrix self-energy for the non-magnetic impurities as,
\begin{eqnarray}
\check{\Sigma} &=& \check{\tau}_3 \cdot \begin{pmatrix}
\Sigma_3 + \mathbf{\Sigma}_3\cdot \boldsymbol{\sigma} & 0 \\
0 & \Sigma_3 + \mathbf{\Sigma}_3\cdot \boldsymbol{\sigma}^T
\end{pmatrix}  + \check{\tau}_0 \cdot \begin{pmatrix}  
\Sigma_0 + \mathbf{\Sigma}\cdot \boldsymbol{\sigma} & 0 \\
0 & \Sigma_0 + \mathbf{\Sigma}\cdot \boldsymbol{\sigma}^T 
\end{pmatrix}, \\
\Sigma_3 &=& \Gamma_u \frac{\cot \delta_s \left[\cot^2 \delta_s- \left(g_0^2+\mathbf{g} \cdot \mathbf{g}\right)\right]}{{D}_{imp}},  \\
\mathbf{\Sigma}_3 &=& \Gamma_u \frac{2\cot \delta_s g_0 \mathbf{g}}{{D}_{imp}}, \\
\Sigma_0 &=&  \Gamma_u  \frac{  g_0\left[\cot^2 \delta_s- \left(g_0^2-\mathbf{g} \cdot \mathbf{g}\right)\right]}{{D}_{imp}},  \\
\mathbf{\Sigma} &=&   \Gamma_u \frac{ \mathbf{g} \left[\cot^2 \delta_s +\left(g_0^2-\mathbf{g} \cdot \mathbf{g}\right)\right]}{{D}_{imp}}, \\
{D}_{imp} &=& \cot^4 \delta_s-2 \cot^2 \delta_s \left(g_0^2 + \mathbf{g}\cdot \mathbf{g}\right)  +  \left(g_0^2 - \mathbf{g}\cdot \mathbf{g}\right)^2,
\label{Eq:TmtSE}
\end{eqnarray}
where $ \delta_s \equiv \tan^{-1}(\pi N_0 V_{imp})$ is the $s$-wave scattering phase shift,  $\Gamma_u = n_{imp}/(\pi N_0)$,  $n_{imp}$ and $V_{imp}$ are the impurity concentration and impurity potential, respectively. . Here $\check{\tau}_3$ is the Pauli matrix in the Nambu space. The $\tau_3$ component of the self-energy that renormalizes the electronic dispersion is ignored. It  can be absorbed in the chemical potential. 
The impurity dressed Green's function reads,
\begin{eqnarray}
\check{\mathbf{G}}^{-1} &=& \check{G}_0^{-1} - \check{\Sigma} = \begin{pmatrix}
\tilde{\omega} -\xi\sigma_0 -\mathbf{\Sigma} \cdot \bsig & -\hat{\Delta} \\
-\hat{\Delta}^\dagger & \tilde{\omega} +\xi\sigma_0 -\mathbf{\Sigma}\cdot \bsig^T
\end{pmatrix}, \nonumber \\
\check{\mathbf{G}}&=&\begin{pmatrix}
\hat{\mathbf{G}}_{11} & \hat{\mathbf{G}}_{12} \\
\hat{\mathbf{G}}_{21} & \hat{\mathbf{G}}_{22}
\end{pmatrix}.
\end{eqnarray}
Here impurity renormalized $\tilde{\omega}=\omega -\Sigma_0$, which is obtained self-consistently. Unlike unitary superconductors, the impurity dressed non-unitary Green's function acquires a different {structure than} the bare Green's function, {in particular the } structure in  spin space for the normal component. The individual component of the  Green's function $\check{\mathbf{G}}$ are,
\begin{eqnarray}
\hat{\mathbf{G}}_{11} &=&  \frac{L_0 + \mathbf{L}_1 \cdot \bsig}{\widetilde{\mathbf{D}}}, \label{G11_dr} \\
\hat{\mathbf{G}}_{22}&=& \frac{L_0(\xi \rightarrow -\xi) + \mathbf{L}_1(\xi \rightarrow -\xi) \cdot \bsig^T}{\widetilde{\mathbf{D}}}, \label{G22_dr} \\
 \hat{\mathbf{G}}_{12} &=& \left[ 2\xi (\bSig \cdot \dv ) + (\rw^2 - \xi^2 - \Delta_0^2 |\dv |^2 + \bSig \cdot \bSig ) \dv \cdot \bsig  + i \Delta_0^2 (\q \times \dv )  \cdot \bsig - 2 (\bSig \cdot \dv ) \bSig \cdot \bsig \right. \nonumber \\
 & & \left. + 2 i \rw (\bSig \times \dv)\cdot \bsig \right] 
 \frac{i\sigma_y \Delta_0}{\widetilde{\mathbf{D}}},  \label{G12_dr} \\
 \hat{\mathbf{G}}_{21} &=&  \left[ -2 \xi  \bSig \cdot \dv^\ast  +  ( \rw^2 -\xi^2- \Delta_0^2 | \dv |^2  + \bSig\cdot\bSig  ) \dv^\ast \cdot \bsig^T    - i \Delta_0^2 (\q \times \dv^\ast)  \cdot \bsig^T -2 \bSig \cdot \dv^\ast \bSig \cdot \bsig^T \right. \nonumber \\
 & & \left.  -2i\rw (\bSig \times \dv^\ast)\cdot \bsig^T \right]  \frac{i \sigma_y \Delta_0}{ \widetilde{\mathbf{D}}},  \label{G21_dr}
\end{eqnarray}
where 
\begin{eqnarray}
L_0 &=& X_0 a_0 - \Delta_0^2 b_0 |\dv |^2 - \Delta_0^2 \q \cdot \bSig, \label{Eq:L0} \\
\mathbf{L}_1 &=& (X_0+ \Delta_0^2 | \dv |^2 ) \bSig + \Delta_0^2 b_0 \q - \Delta_0^2 \bSig \cdot \dv \dv^\ast  - \Delta_0^2 \bSig \cdot \dv^\ast \dv,  \label{Eq:L1} 
\end{eqnarray}
Here $a_0/b_0=\rw\mp \xi$,  $X_0 = b_0^2-\bSig\cdot \bSig$ 
and the denominator $\widetilde{\mathbf{D}} = (\xi^2 + \rQ_+^2) ( \xi^2 + \rQ_-^2)$, where $\rQ_\pm$ is,
\begin{eqnarray}
\rQ_\pm^2 &=& \Delta_0^2 | \dv |^2 -\rw^2 - \bSig \cdot \bSig  \pm \sqrt{\Delta_0^4 \q \cdot \q + 4\rw^2 \bSig \cdot \bSig + 4\Delta_0^2 \rw \q \cdot \bSig -  4 \Delta_0^2   (\bSig \cdot \dv) (\bSig \cdot \dv^\ast) }
\end{eqnarray}
For the unitary states, the nodes are symmetry imposed, and for the non-unitary states, nodes may shift away from the high symmetry directions, and their positions remain protected against disorder as long as the $\bSig$ component of the impurity self-energy vanishes, $\bSig$ can be interpreted as impurity induced magnetization. For the chiral states, this term remains finite and give rise to non-degenerate quasiparticle spin density, and in principle, can    change the nodal positions. Here the nodes do not refer to the zeros of gap or the order parameter, instead, they are the zeros in the quasi-particle spectrum on the Fermi surface. In the unitary states, the gap nodes and the spectral nodes are same unlike the nonunitary states.  There are some additional triplet terms $\bSig \cdot \bsig$ and $(\bSig \times \dv)\cdot \bsig$ in Eq. \eqref{G12_dr}, these terms reflect impurity induced modification of spin structure of the Cooper pairs.   It is worth mentioning that there is an impurity induced odd-frequency pairing for the chiral non-unitary states, which is  spin singlet and odd parity in nature. 

After obtaining the impurity dressed Green's function, we calculate the electronic thermal conductivity $\kappa$ using the Kubo formula that relates the thermal conductivity to the heat-current response\cite{AmbegaokarGriffin:1994}. We ignore the vertex corrections and restrict ourselves to the bare thermal current response function. The vertex corrections are small in the strong scattering limit that we focus on this article\citep{PJH_UPt3_1995}. The diagonal thermal conductivity for a general triplet superconductor reads,
\begin{equation}
\frac{\kappa_{ii}}{T} =  \int_{-\infty}^{\infty} d\omega \frac{\omega^2}{T^2} \left(- \frac{d n_{F} (\omega) }{d\omega}\right) \left\langle  N_0 v_{Fi}^2 \frac{ (-c_1 c_4 + c_2 c_3 )+ c_3 b_1 + c_1 b_2 +  b_3 ( -c_3 +c_1 c_2 )/c_4}{ ( c_3^2 + c_1^2 c_4 - c_1 c_2 c_3)} \right\rangle_{FS}.
\label{Eq:kappa_f}
\end{equation}
Here $c_1 = -2\mathrm{Re}(\rQ_+ + \rQ_-) $, $c_2 =|\rQ_+|^2 + |\rQ_-|^2 + 4 \mathrm{Re}(\rQ_+)  \mathrm{Re}(\rQ_-)$, $c_3 =  -2 |\rQ_+|^2  \mathrm{Re}(\rQ_-) -  2 |\rQ_-|^2  \mathrm{Re}(\rQ_+)$, $c_4 = |\rQ_+|^2 |\rQ_-|^2$
and the coefficients $b_{i=1,..,3}$ are,
\begin{eqnarray}
b_1 &=&  (| \rw|^2 -\Delta_0^2 |\dv |^2 + | \bSig |^2 ) + \mathrm{Re}\left[ \tilde{Q}_+^2 + \tilde{Q}^{2}_- \right] \\
b_2&=&\frac{1}{4} | \tilde{Q}_+^2 + \tilde{Q}^{2}_- |^2 + (|\rw |^2 + |\bSig |^2 - \Delta_0^2 |\dv |^2) \mathrm{Re}\left[ \tilde{Q}_+^2 + \tilde{Q}^{2}_- \right] +  3\Delta_0^4 \q \cdot \q + 4 |\bSig |^2 ( |\rw |^2 +\mathrm{Re}[\rw^2] )  \nonumber \\
& & - 4\Delta_0^2 \left( | \bSig \cdot \dv |^2 + |\bSig \cdot \dv^\ast |^2 + \mathrm{Re}[(\bSig \cdot \dv) (\bSig \cdot \dv^\ast) ] \right) \nonumber \\
& &+ 4 |\rw |^2 \mathrm{Re}[ \bSig \cdot \bSig] + 4\Delta_0^2 \mathrm{Re}[ \rw \q \cdot \bSig^\ast  + 2 \rw \q \cdot \bSig] \\
b_3 &=&  ( |\rw |^2 -\Delta_0^2 |\dv |^2 ) \left[ |\alpha_+ |^2 + \Delta_0^4 \q \cdot \q \right]- 2 \Delta_0^4 \q \cdot \q \mathrm{Re}[\alpha_+] + \mathcal{Y}(\bSig) \\
 {\mathcal{Y}}(\bSig)  &=&  \left\lbrace |(\rw^2 + \Delta_0^2 |\dv |^2 - \bSig \cdot \bSig)  |^2 -\Delta_0^4 \q \cdot \q   \right\rbrace |\bSig |^2 - 2\Delta_0^2 \mathrm{Re}[ (\alpha_- - \alpha_-^\ast ) \rw \q\cdot \bSig^\ast ] \nonumber \\
 & & - 4\Delta_0^2 (|\rw |^2+ |\bSig |^2 -\Delta_0^2 |\dv |^2 )  \mathrm{Re}[ \rw \q \cdot \bSig^\ast ] +  2 \Delta_0^4 \mathrm{Re}\left[| \q \cdot \bSig |^2 -(\q \cdot \bSig )^2+ \bSig \cdot \bSig \q \cdot \q \right] \nonumber \\
 && + 2\Delta_0^2\mathrm{Re} \left[(  | \rw |^2- \rw^2 ) - ( |\bSig |^2 -\bSig \cdot \bSig ) \right] ( |\bSig \cdot \dv |^2 + |\bSig \cdot \dv^\ast |^2 ) -4 \Delta_0^2 |\dv |^2 |\rw |^2 | \bSig |^2  \nonumber \\
 & &  +4\Delta_0^2 \mathrm{Re}\left[ \alpha_+ (\bSig^\ast \cdot \dv^\ast ) (\bSig^\ast \cdot \dv ) \right] - 4\Delta_0^2 \mathrm{Re}[ (\rw^2 - \Delta_0^2 |\dv |^2 ) \bSig^\ast \cdot \bSig^\ast ],
\end{eqnarray}
where $\alpha_\pm = \rw^2 -\Delta_0^2 |\dv |^2 \pm \bSig \cdot \bSig$. The derivation of thermal conductivity is provided in the supplementary materials. We calculate the full temperature dependence of thermal conductivity using self-consistently determined superconducting gap using an effective paring potential to give a single transition temperature (see appendix ). 
\section{Results \& Discussion}
\begin{figure}
\includegraphics[width=1\columnwidth]{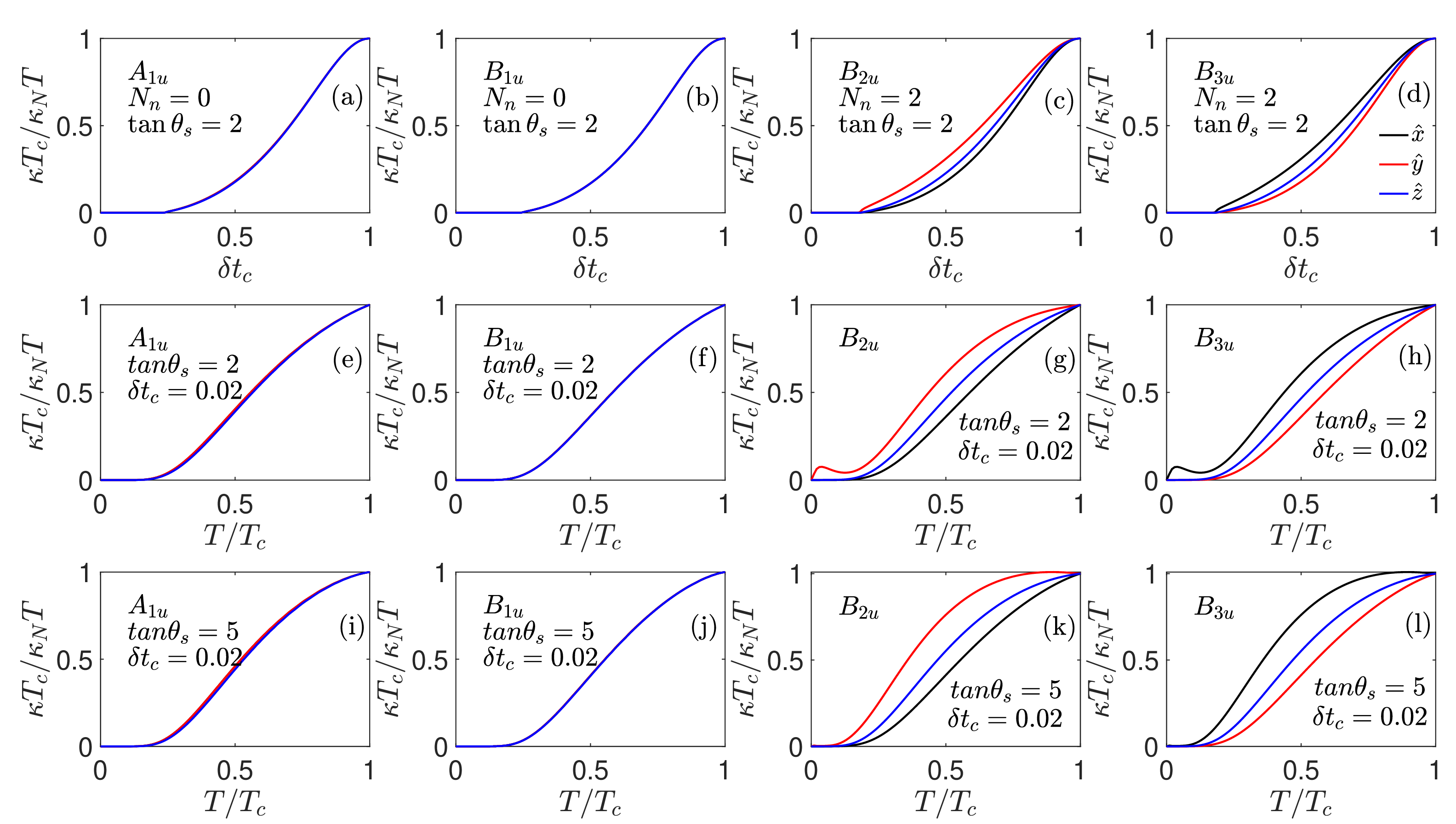} 
\caption{{\bf Thermal conductivity for the single component unitary states allowed by the \d2h point group:}  The thermal conductivity normalized to its value at $T_c$ for the four irreducible representations of the \d2h point group shown column-wise   for the $A_{1u}$, $B_{1u}$, $B_{2u}$ and $B_{3u}$ representations from panel (a) to (d), respectively. The first row shows the residual thermal conductivity in the zero temperature limit as a function of relative reduction in the transition temperature $\delta t_c$. Panels (e) to (h) in the second row show  $\kappa T_c/\kappa_N T$ for the weak scatterers with $\tan \theta_s =2$ and for the intermediate strength scatterers with  $\tan \theta_s =5$   from  panels (i) to(l) in the third row. For the temperature dependence, the $T_c$ is reduced by $2\% (\delta t_c=0.02)$ \textit{w.r.t.} the clean limit.  } 
\label{fig:1u} 
\end{figure}
\subsection{$T=0$ limit of density of states and thermal conductivity}
We start with the single component states based on four irreducible representations of the \d2h point group symmetry. The basis functions for these four states are listed in Table \ref{Table1},  where the  $A_{1u}$ state remains gapped unless the coefficient of one of the basis functions is set to zero. We exclude that possibility and choose the same coefficients for all three basis functions, and this choice of coefficients is adopted for the other states, too. For qualitative understanding of the low energy properties, this is a reasonable choice. In principle, it is also possible to generate line nodes with appropriate choice of basis function coefficients, but those possibilities are excluded considering the recent experimental results on \ute2. Apart from the $A_{1u}$ state, the $B_{1u}$ state also remains gapped because the open Fermi surface along the $\hat{z}$-axis forbids the nodes for this state. For the $B_{2u}$ and $B_{3u}$ states, there is a pair of point nodes that exists along the $\hat{y}$-axis and $\hat{x}$-axis, respectively. Fig. \ref{fig:1u}(a) to (d) show the disorder dependence of the thermal conductivity normalized to the normal state value at the transition as a function of relative reduction in the transition temperature $\delta t_c \equiv = 1-T_c/T_{c0}$, where $T_{c0}$ is the transition temperature in the clean limit. For a clean system, $\delta t_c=0$ and $\delta t_c$ reaches unity as the impurity scattering kill superconductivity. The normalized residual thermal conductivity $\kappa T_c/\kappa_N T$ in the zero temperature limit remains zero up to a threshold value of disorder $\Gamma_{th}$ for all four states, this $\Gamma_{th}$ corresponds to a threshold level of $T_c$ suppression $\delta t_c^{th}$. This threshold value of disorder depends on the superconducting gap structure and the strength of the impurity potential\cite{PJHirschfeld:1986}. For the gapped states $A_{1u}$ and $B_{1u}$, the residual thermal conductivity remains zero for slightly higher values of disorder compared to the other two states $B_{2u}$ and $B_{3u}$ as expected due to the presence of impurity induced quasiparticle states near the nodes. The normalized thermal conductivity remains very isotropic for the $A_{1u}$ and $B_{1u}$ states. However, for the nodal states, the residual thermal conductivity is enhanced for thermal current along the nodal directions. This trend in anisotropy in thermal conductivity continues even at finite temperatures, as shown in the temperature evolution of $\kappa T_c/\kappa_N T$ for the four single component states in Fig. \ref{fig:1u}(e) to (h) for $\tan \theta_s =2$ and in panels (i) to (l) for $\tan \theta_s =5$, which represents the stronger scatterers. For both impurity strengths, $\delta t_c$ is $0.02$. In the presence of the point nodes, the thermal conductivity shows a weak maximum  at very low temperature along the nodal direction for the weak scatterers, which disappears as the scattering rate increases. This is a known  behavior for the  superconducting states with point nodes\cite{PJHirschfeld:1988}, which  is not present for the stronger impurity potentials.  

\begin{figure*}
\includegraphics[width=.49\columnwidth]{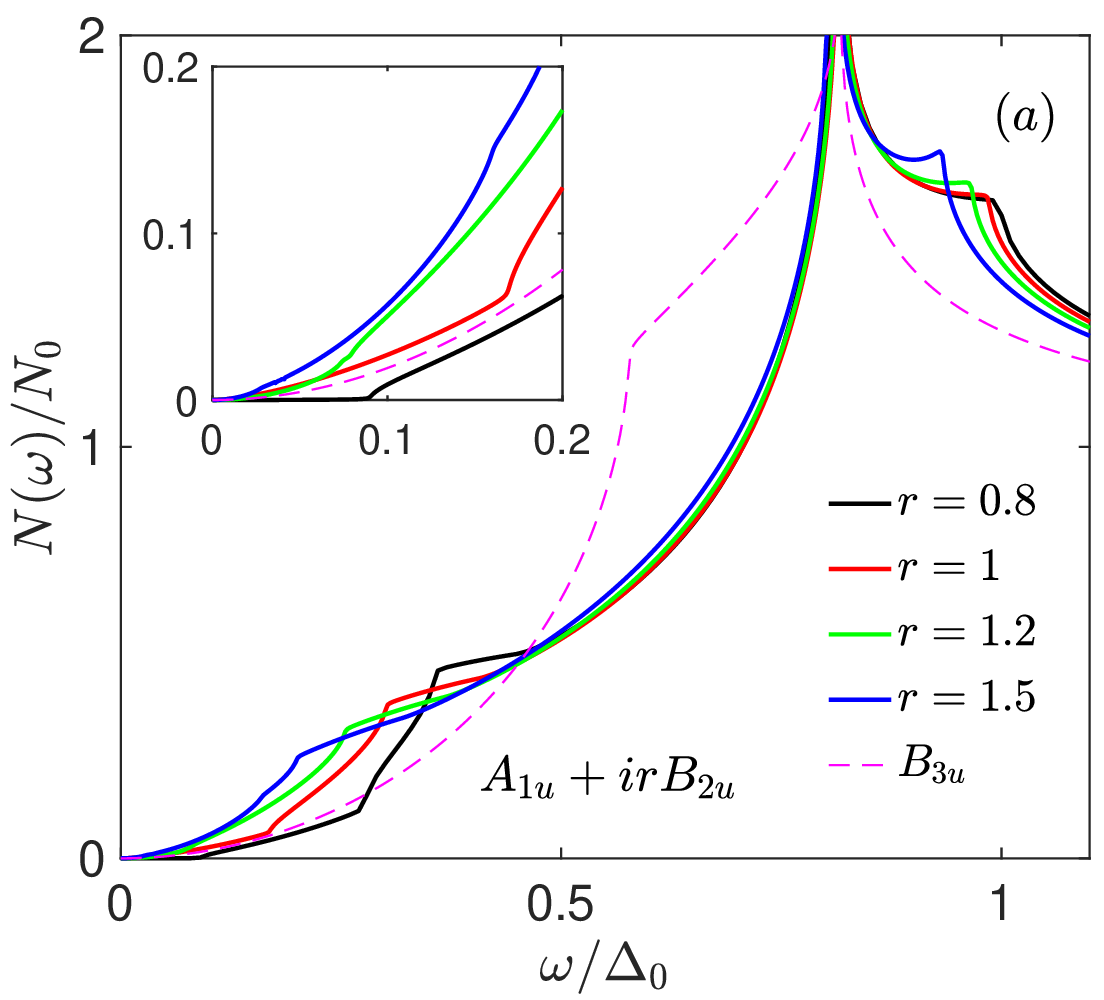} 
\includegraphics[width=.49\columnwidth]{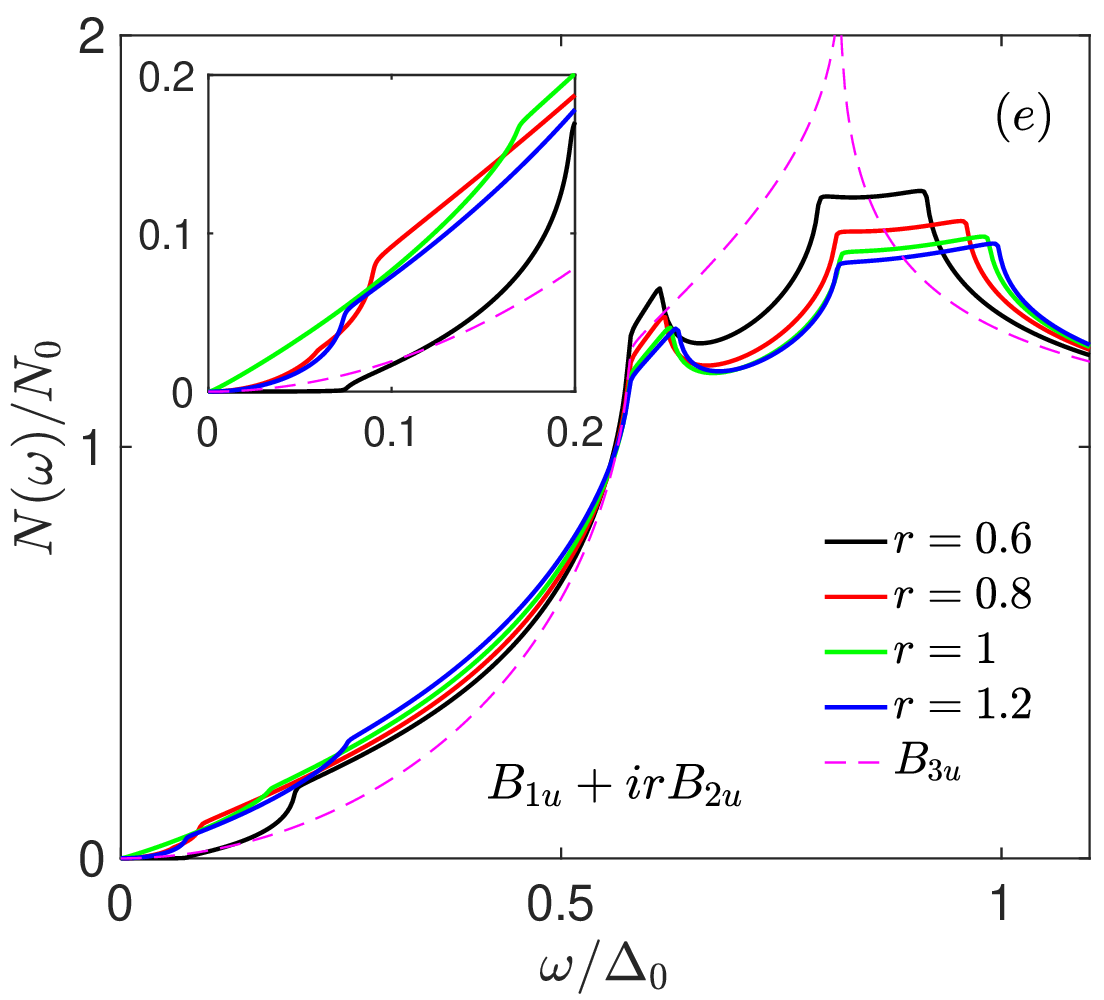} 
\includegraphics[width=.49\columnwidth]{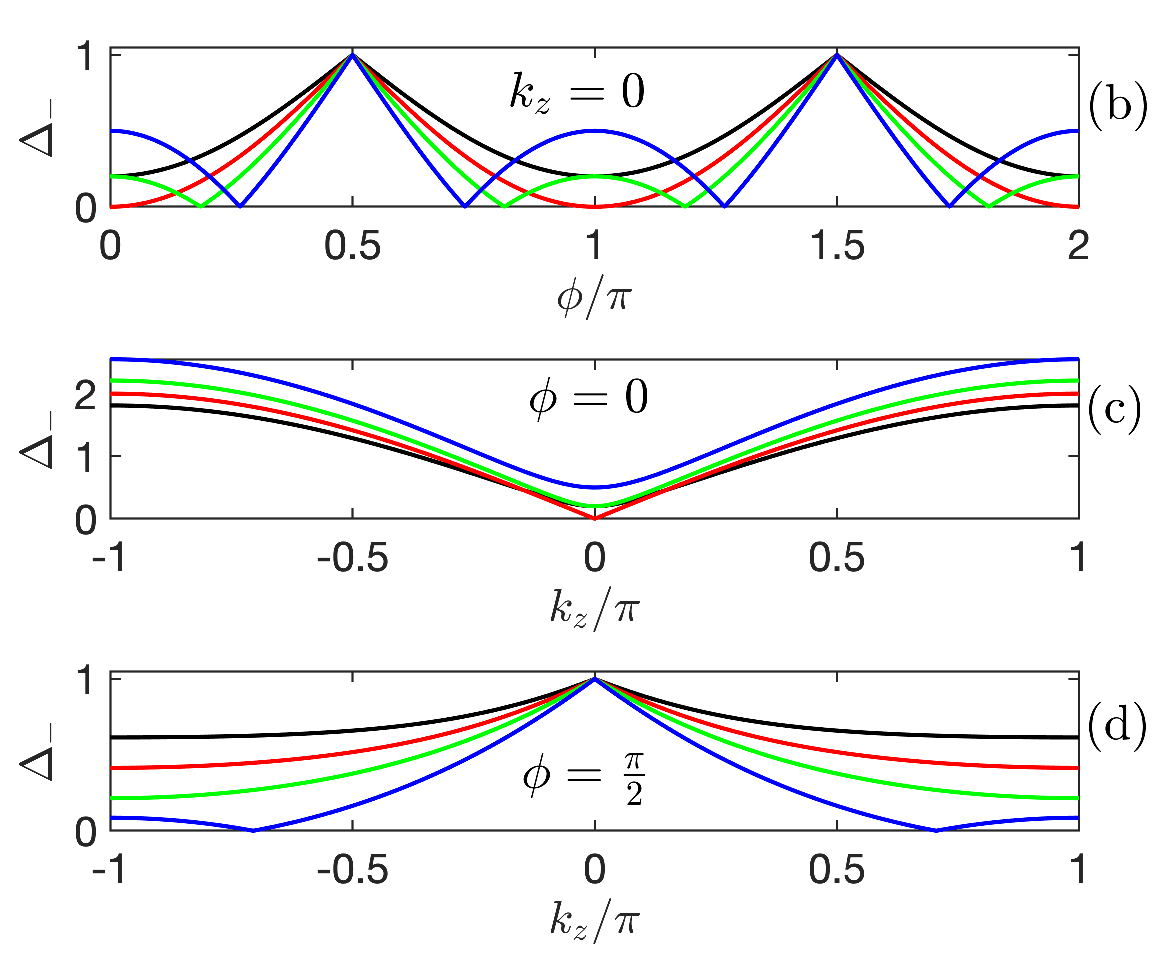} 
\includegraphics[width=.49\columnwidth]{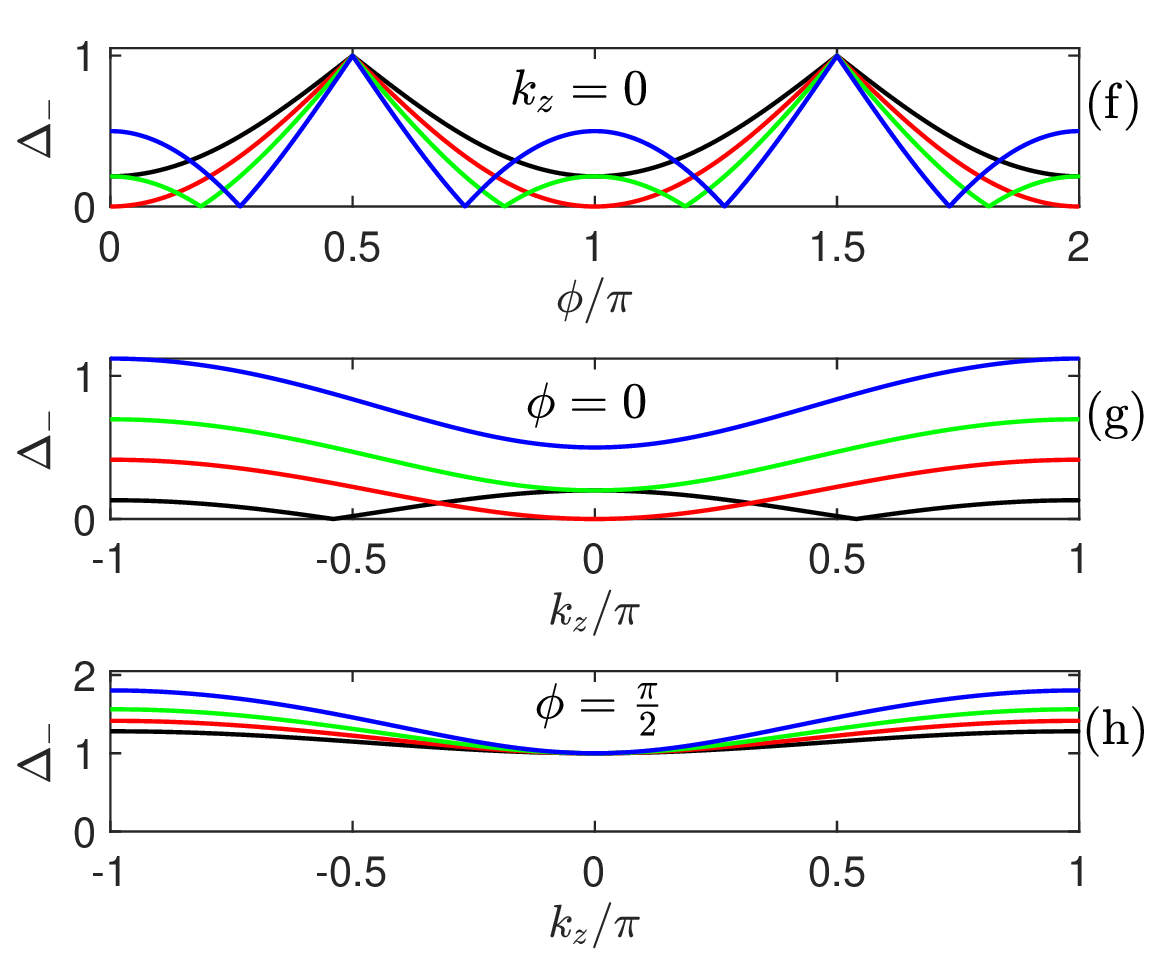} 
\caption{{\bf Density of states:} Panel (a) shows the DOS per spin for the $A_{1u}+irB_{2u}$ state for different values of the mixing parameter $r$. The inset shows the energy dependence of DOS at low energies. A dashed line illustrating  the DOS for the unitary $B_{3u}$ state has been shown for comparison.  Panel (b), (c) and (d) show the variation of $\Delta_-$ for different values of $r$ in the $xy$, $xz$ and $yz$ planes, respectively. Panel (f) shows the DOS per spin for the  $B_{1u}+irB_{2u}$ state and $\Delta_-$  in the $x{}y$, $x{}z$ and $y{}z$ planes are shown in panels (f), (g) and (h), respectively. } 
\label{fig:DOS1} 
\end{figure*}

It is useful to examine the density of states (DOS) and the  structure of the low energy quasiparticle states before discussing the thermal transport for the nonunitary states.  We first report the average density of states per spin  for the nonunitary states on a cylindrical Fermi surface that is open along the $\hat{z}$-axis. Fig. \ref{fig:DOS1}(a) shows the DOS for the $A_{1u}+irB_{2u}$ state, which is a chiral state, and Fig. \ref{fig:DOS1}(b), (c) and (d) show $\Delta_-$ for this state. In the gapped phase ($r<1$), this state has  minima near the $\hat{x}$-axis and a small gap is visible in the DOS. For $r=1$, a pair of point nodes appears along the $\hat{x}$-axis and the low energy DOS shows $\omega^2$ behavior that is expected for linear point nodes. Here and in subsequent discussion, ``node" refers to spectral nodes, not the gap nodes. However, in contrast to a unitary state, the low energy quadratic behavior remains confined to very low energy scale as compared to the unitary $B_{3u}$ state.   For $r>1$, the nodes move away from the $\hat{x}$-axis and the positions of four nodes are  determined by $\tan \phi =\pm \sqrt{r^2-1}$, where $\phi$ is the polar angle on the cylindrical Fermi surface. As the mixing parameter $r$ increases, additional pairs of nodes appear in the $y{}z$ plane at $\sin (k_z/2)=\pm 1/\sqrt{r^2-1}$. In the $r\rightarrow \infty$ limit, only two point nodes along $\hat{y}$-axis survive, as expected for a pure $B_{2u}$ state. The low energy DOS remains quadratic in all these cases. For $1\le r < \sqrt{2}$, the nodes remain closer to the $\hat{x}$-axis, for $r\ge \sqrt{2}$, the nodes move closer to $\hat{y}$-axis. The $A_{1u}+irB_{3u}$ state also shows similar DOS as $A_{1u}+irB_{2u}$ state, but it has different nodal structure. It has gap  minima in quasiparticle spectrum near the $\hat{y}$-axis, and  nodes appear along the $\hat{y}$-axis. For $r>1$, a set of four nodes appear near $\hat{y}$-axis at $\phi=\pm \cot^{-1} \sqrt{r^2-1}$ and move towards the $\hat{x}$-axis in the $x{}y$ plane as the value of $r$ increases. For $r>\sqrt{2}$, four more nodes appear at $k_z=\pm 2\sin^{-1}(1/\sqrt{r^2-1})$ in the $x{}z$ plane. Both these states are chiral and show finite quasiparticle spin density along $\hat{y}$ and $\hat{x}$ directions in the spin space. The last AF non-unitary state is $A_{1u}+irB_{1u}$, which is not a chiral state.   It has nodes  along the $\hat{x}$ and $\hat{y}$ directions for $r=1$, a quadruple pair of point nodes in the $xz$-plane, and another set of four point nodes $yz$ plane, where the $k_z$ for the nodal position is determined by $\sin (k_z/2)=\pm\sqrt{r^2-1} $. The DOS shows quadratic behavior at low energies (see  appendix). 

Next, we consider the ferromagnetic nonunitary states on the cylindrical Fermi surface, which are chiral states with finite {Cooper pair} spin moment. Fig. \ref{fig:DOS1}(d) shows the DOS for the $B_{1u}+irB_{2u}$ state along with the $\Delta_-$ in the panels (e) and (f). This state is gapped  for $r< 1/\sqrt{2}$, and  a quadruple pair of point nodes appears in the $y{}z$ plane for $1/\sqrt{2} \le r < 1$, whose positions are determined  by $\sin (k_z/2) = r/\sqrt{1-r^2}$. For $r>1$, a quadruple pair of point nodes appear in the $x{}y$ plane  at $\phi = \pm \tan^{-1} \sqrt{r^2-1}$  close to the $\hat{x}$-axis and move towards $\hat{y}$-axis as $r$ increases. These states show quadratic DOS at low energies. For $r=1$, this state shows point nodes along the $\hat{x}$-axis, but these are the quadratic point nodes. The first derivative of the $\Delta_-$ vanishes at the nodes for a quadratic or second order point nodes. This leads to linear DOS at low energies, as shown in the Fig. \ref{fig:DOS1}(d). Similarly, for the $B_{1u}+irB_{3u}$ a twin pair of quadratic point nodes appear along the $\hat{y}$-axis and shows linear DOS at low energies for $r=1$. For $1/\sqrt{2} \le r < 1$ four point nodes appear at $\sin (k_z/2)=\pm \sqrt{1-r^2}/r$ in the  $y{}z$ plane,  and for $r>1$ there is a  quadruple pair of point nodes  in the  $x{}y$ plane at $\cot \phi = \pm \sqrt{r^2-1}$, closer to the $\hat{y}$-axis for $r\geq 1$, which move towards the $\hat{x}$-axis for $r\gg 1$. The DOS remains quadratic, as expected. For the $B_{2u}+irB_{3u}$ state, four point nodes are either located in the $y{}z$ plane at $\sin(k_z/2)=\pm r/\sqrt{1-r^2}$ for $r\leq1/\sqrt{2}$ or in the $x{}z$ plane at $\sin(k_z/2)=\pm 1/\sqrt{r^2-1}$ for $r\geq \sqrt{2}$. These states show $\omega^2$ behavior in the low energy DOS. For $1/\sqrt{2}< r < \sqrt{2}$, there exists a gap in the quasiparticle spectrum due to lack of nodes (see supplementary materials). In contrast to the AF nonunitary states on the cylindrical Fermi surface, the FM nonunitary states can have at most four nodes and are expected to be more anisotropic. 

\begin{figure}
\includegraphics[width=.49\columnwidth]{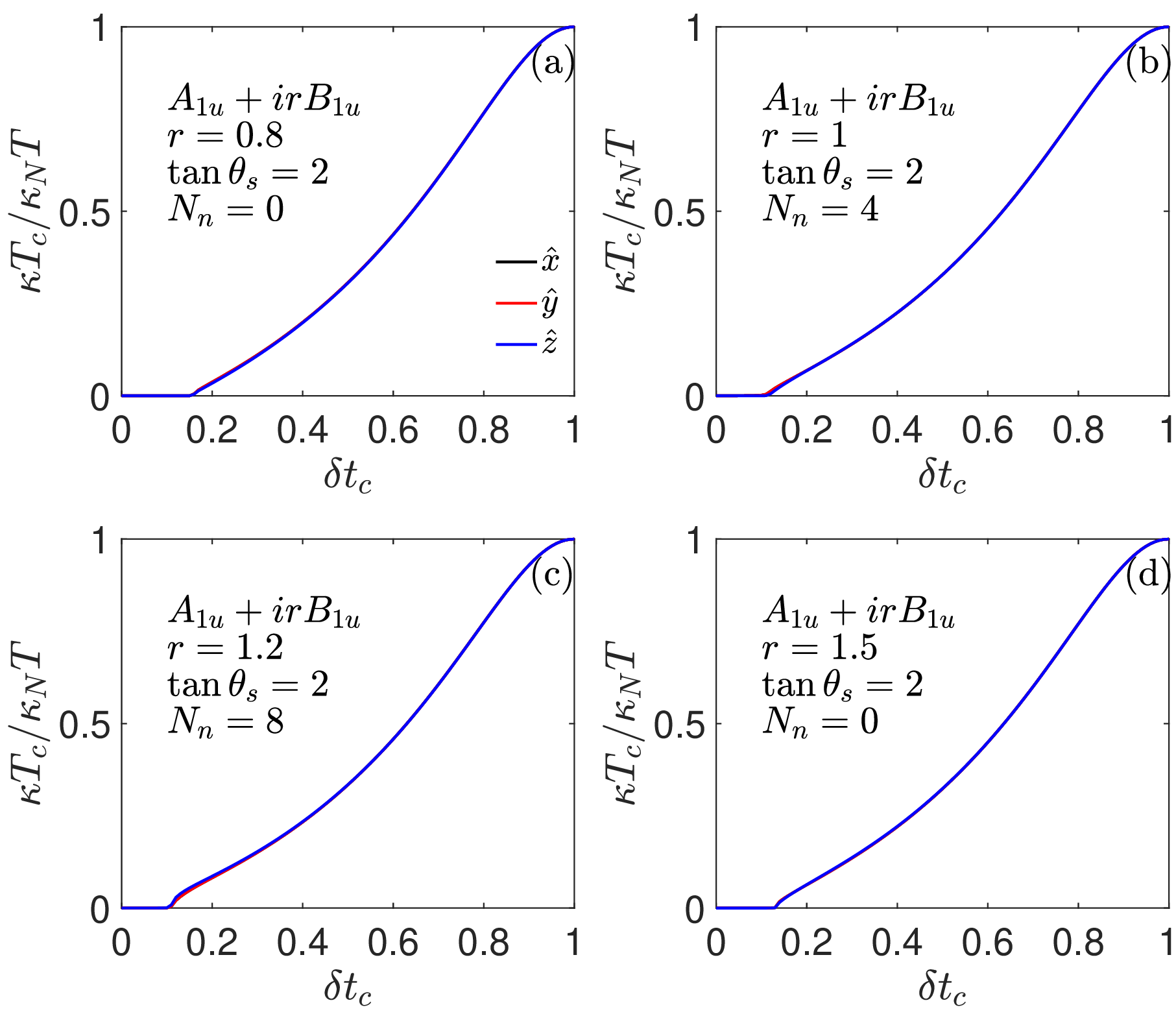} 
\includegraphics[width=.49\columnwidth]{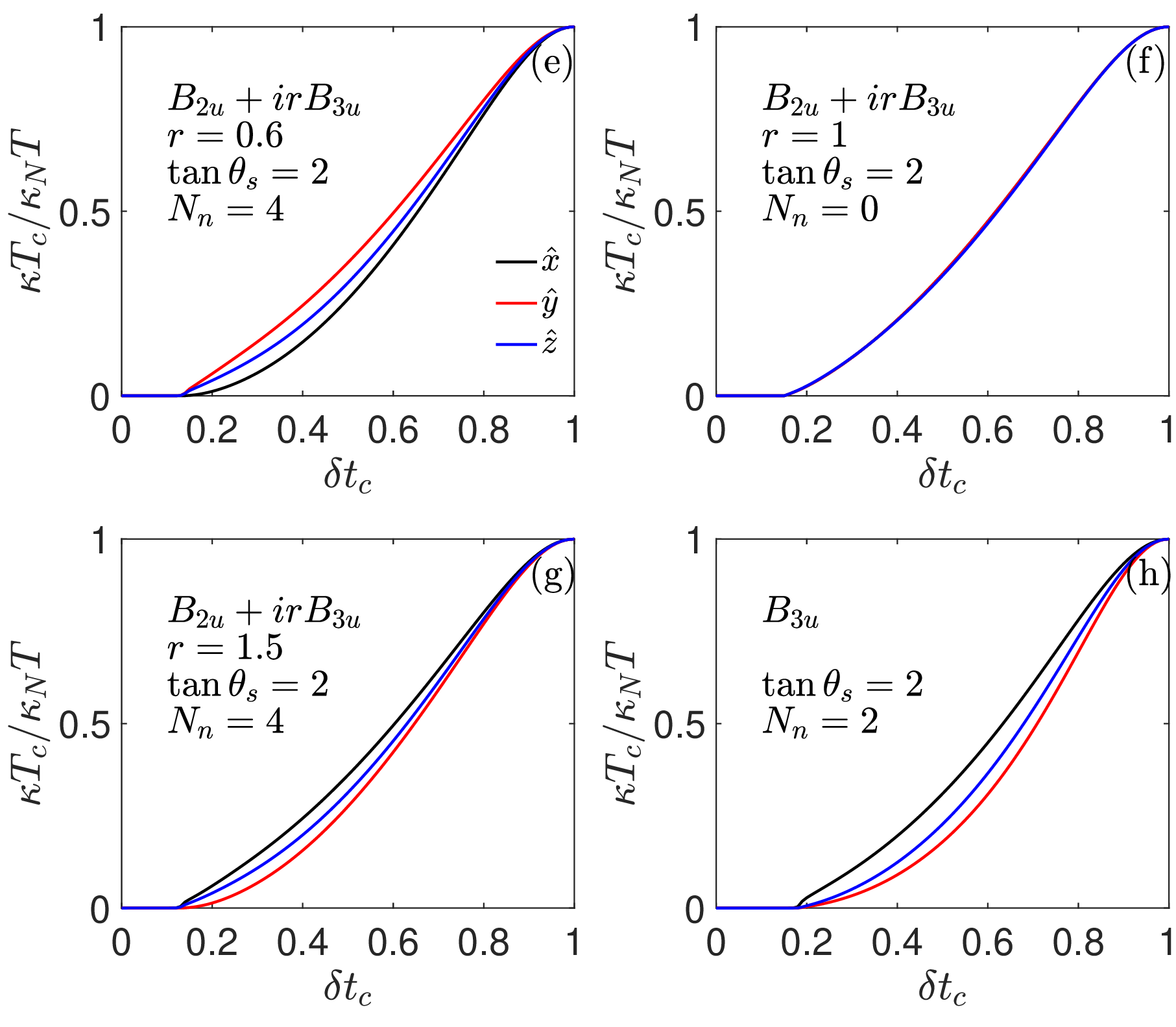} 
\caption{{\bf Residual thermal conductivity:}  Zero temperature limit $\kappa T_c/\kappa_N T$ for the $A_{1u}+irB_{1u}$ in panels (a) to (d) and for the $B_{2u}+irB_{3u}$    state in panels (e) to (g) for various values of mixing parameter $r$. Panel (h) shows $\kappa T_c/\kappa_N T|_{T\rightarrow 0}$ for the unitary $B_{3u}$ state. The $s$-wave scattering phase shift is $\theta_s =\tan^{-1}(2)$ for all the panels. The number of nodes $N_n$ is indicated for each case.  } 
\label{fig:k01} 
\end{figure}

We now discuss the zero temperature limit of the thermal conductivity, which is very sensitive to the gap structure.  We compare the normalized $\kappa T_c/T \kappa_N$ along three principle directions, where  the thermal conductivity along a particular direction is normalized to its normal state value at $T_c$ along that direction. This suppresses the intrinsic anisotropy present in the electronic structure {and highlights the effect of order parameter anisotropy}. Panels (a) to (d) in Fig. \ref{fig:k01} show the $\kappa T_c/T \kappa_N$ in the zero temperature limit for the $A_{1u}+irB_{1u}$ state as a function of relative reduction in the transition temperature $\delta T_c$.  For a weakly disordered system, $\kappa/T$ vanishes in the zero temperature limit, but as the disorder level increases and crosses $\Gamma_{th}$, $\kappa/T|_{T\rightarrow 0}$ becomes finite and reaches the normal state value as the superconductivity vanishes. For the $A_{1u}+irB_{1u}$ state, the $\kappa T_c/\kappa_N T$ shows  isotropic behavior in the $x{}y$-plane. Note, for this state, the nodes are always along the $\hat{x}$ and $\hat{y}$ axes at the same $k_z$ value. In contrast, the $B_{2u}+irB_{3u}$ state shows    relatively weaker level anisotropy than a unitary state. Panels (e) to (g) in Fig. \ref{fig:k01} show the $\kappa T_c/\kappa_N T$ for the $B_{2u}+irB_{3u}$  state, while Fig. \ref{fig:k01}(h) shows $\kappa T_c/\kappa_N T$  for the unitary $B_{3u}$ state with nodes along $\hat{x}$ axis. As shown in Fig. \ref{fig:k01}(e), with nodes in the $y{}z$-plane $\kappa_{yy}$ and $\kappa_{zz}$ exceed $\kappa_{xx}$, and this behavior reverses as the nodes move to the $x{}z$-plane for the $B_{2u}+irB_{3u}$ state, as depicted in  Fig. \ref{fig:k01}(h). As the value of $r$ increases, the anisotropy also reduces, and for $r=1$ the superconducting state becomes fully isotropic, as shown in Fig. \ref{fig:k01}(f). This is a special case, which has a four-fold symmetric $\Delta_-$, leads to a very isotropic normalized thermal conductivity along the three principal directions. As $r$ becomes larger than unity, the $\kappa_{xx}$ starts to dominate.
\begin{figure*}
\includegraphics[width=.49\columnwidth]{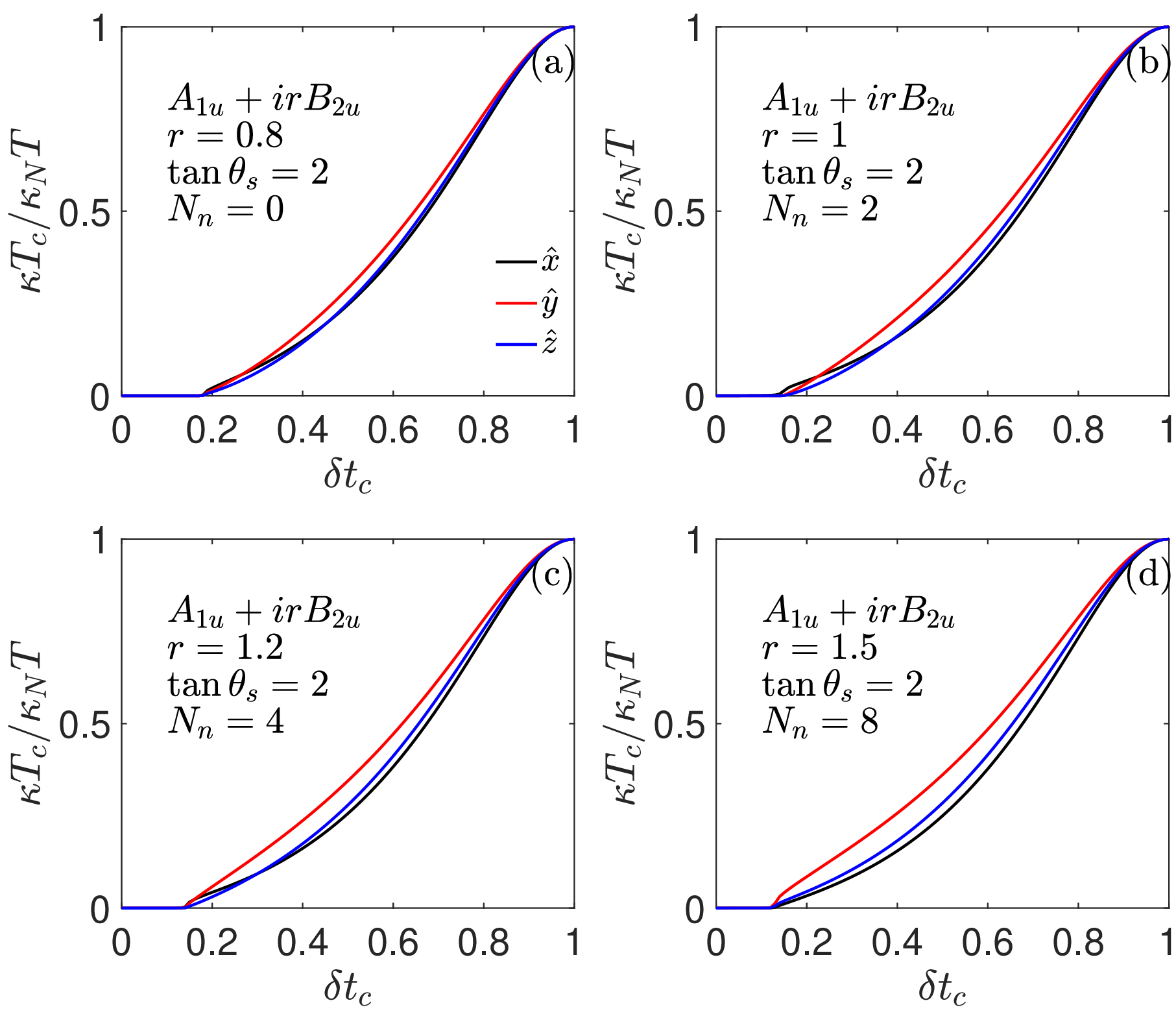} 
\includegraphics[width=.49\columnwidth]{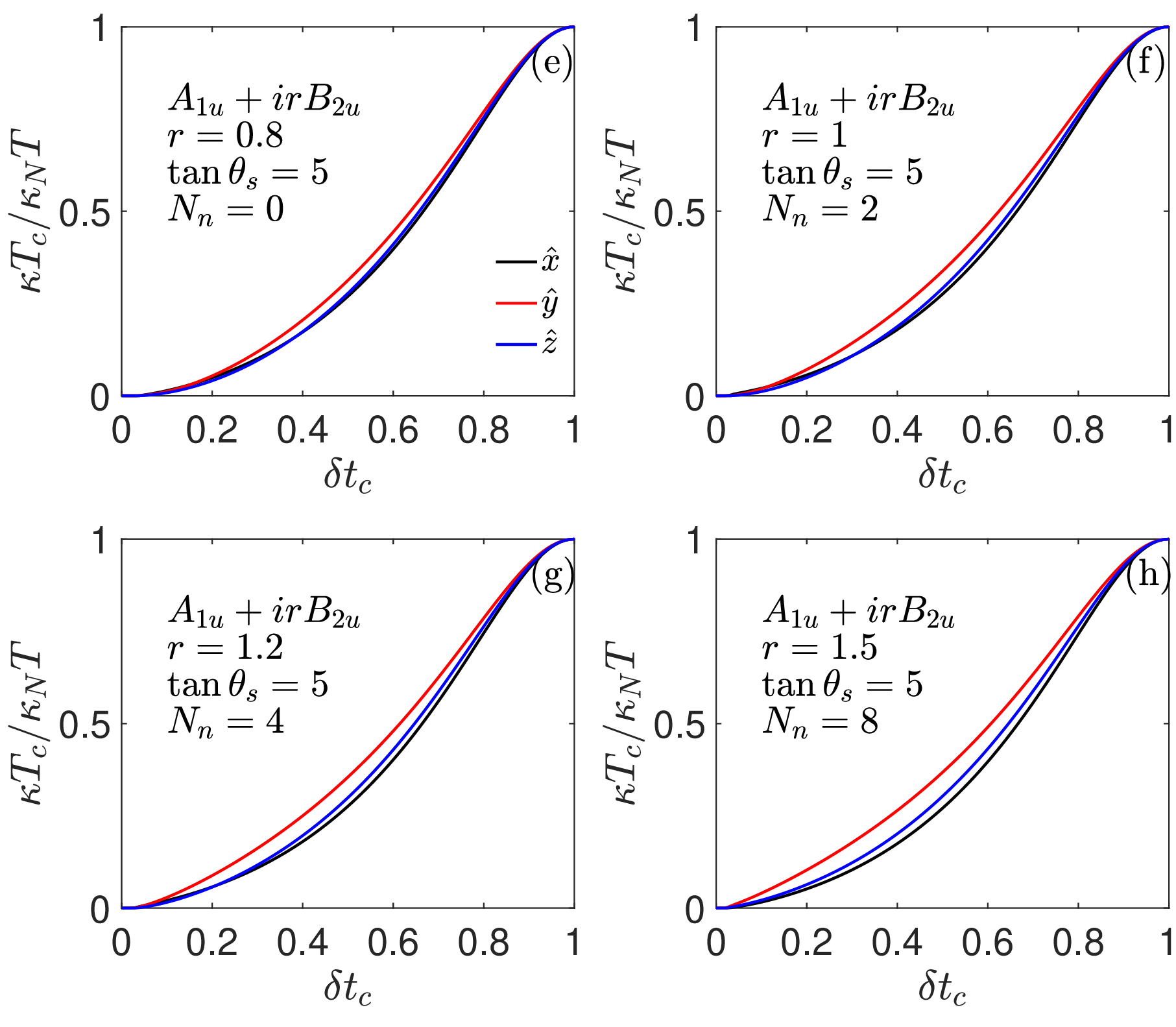} 
\includegraphics[width=.49\columnwidth]{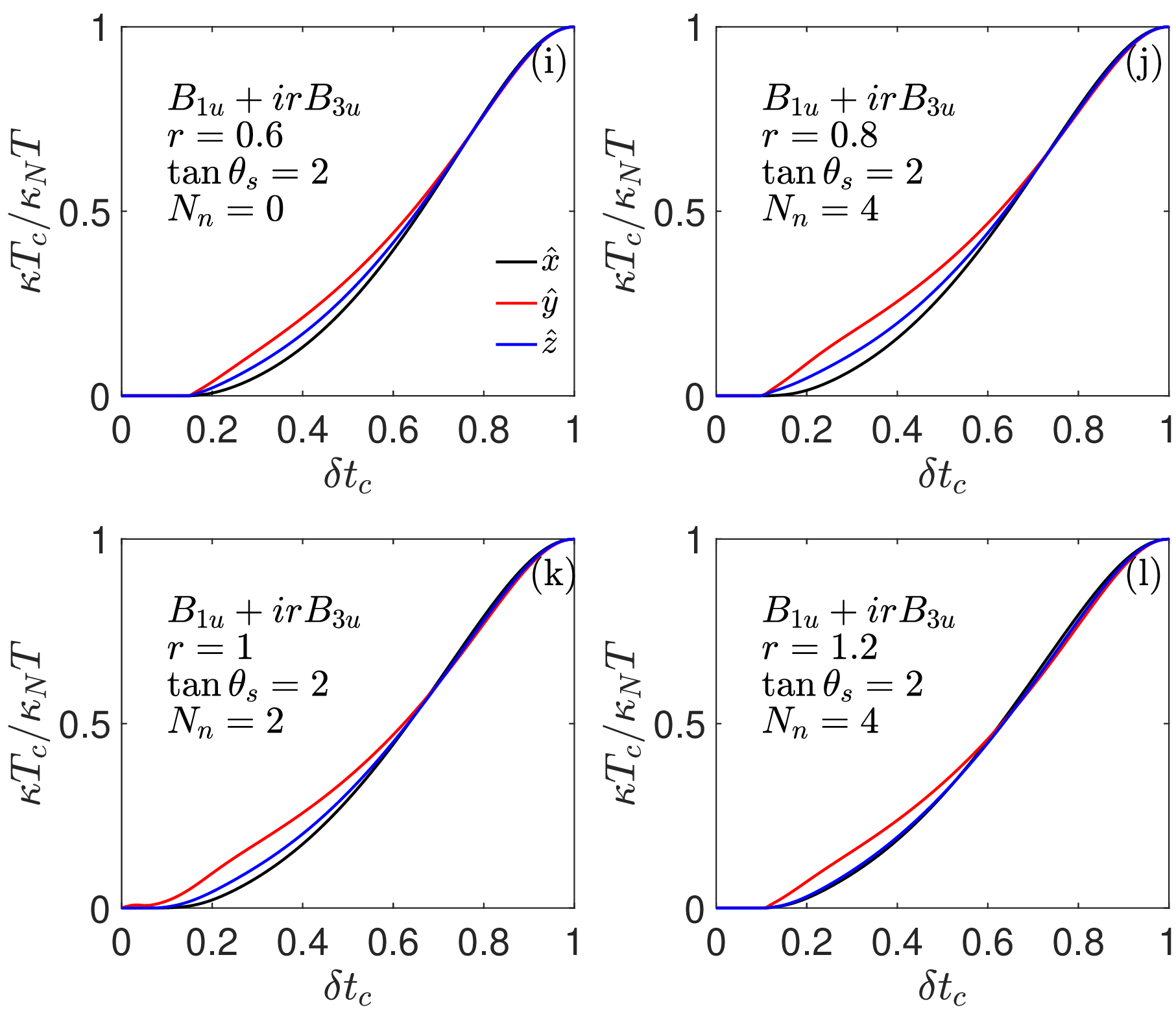} 
\includegraphics[width=.49\columnwidth]{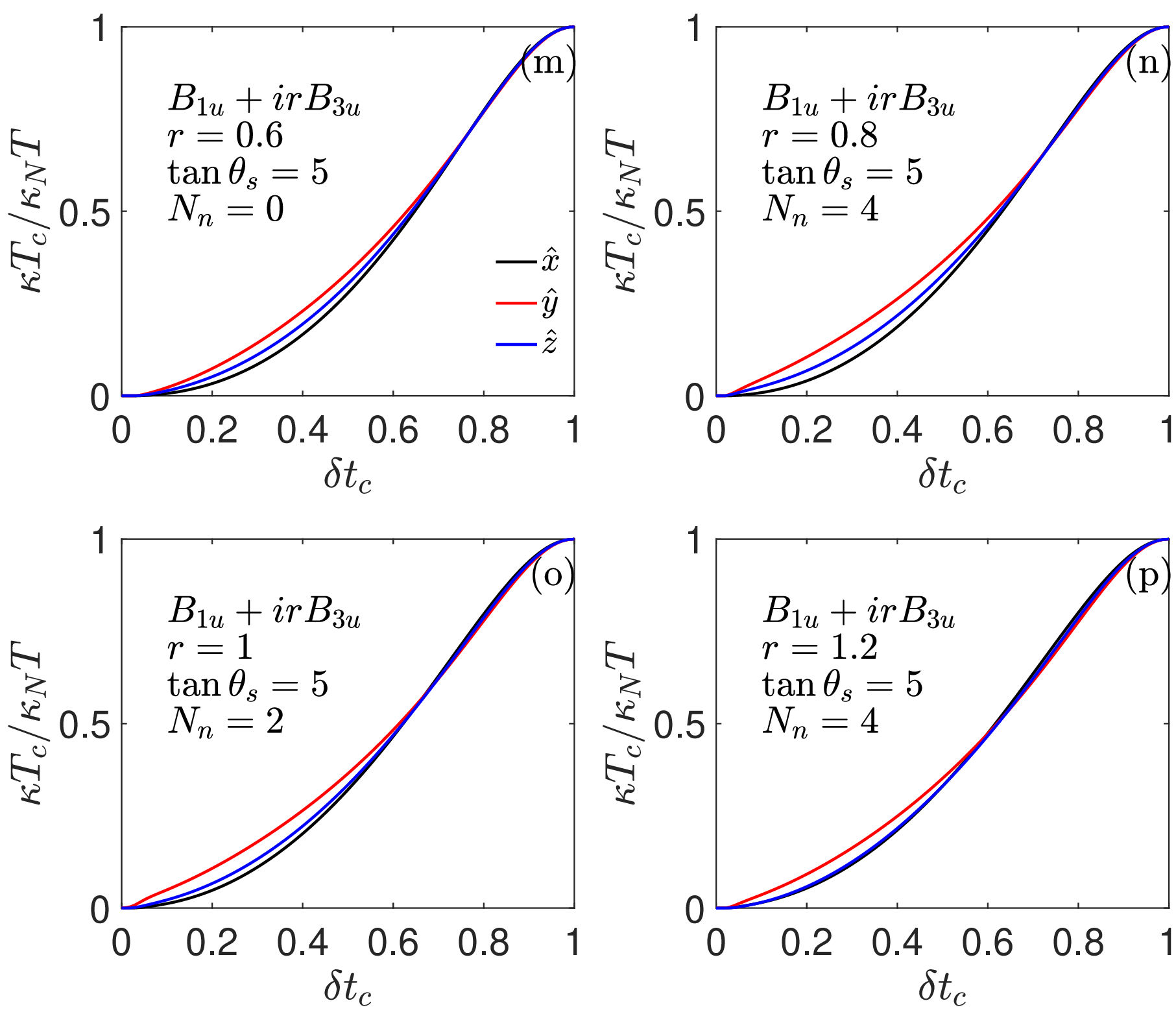} 
\caption{{\bf Residual thermal conductivity:} Panels (a) to (h) show $\kappa/T$ in the zero temperature limit normalized to $\kappa_N/T$ at $T_c$ as a function of relative  $T_c$ reduction $\delta t_c$  for the $A_{1u}+irB_{2u}$ state for different values of the mixing parameter $r$. The $s$-wave scattering phase for the panels (a) to (d) is $\theta_s =\tan^{-1}(2)$ and for the panels (e) to (h) is $\theta_s = \tan^{-1} (5)$.   For the  $B_{2u}+irB_{3u}$ state, $\kappa T_c/\kappa_N T|_{T\rightarrow 0}$  is shown for various values of parameter $r$ for $s$-wave scattering phase shift $\theta_s =\tan^{-1}(2)$ in the panels (i) to (l) and with $\theta_s =\tan^{-1}(5)$ in the panels (m) to (p). The number of nodes $N_n$ is indicated for each case.  } 
\label{fig:k02} 
\end{figure*}

Next, we look at the zero temperature limit thermal conductivity for the $A_{1u}+irB_{2u}$ state for two different impurity potential strengths. Fig. \ref{fig:k02} (a) to (d) show $\kappa T_c/\kappa_N T$ for $s$-wave scattering phase shift $\tan^{-1}(2)$ and panels (e) to (h) show $\kappa T_c/\kappa_N T$ for $\theta_s=\tan^{-1}(5)$. The 
$\kappa/T|_{T\rightarrow 0}$  becomes finite above a threshold disorder level, like earlier cases. This threshold scattering rate is smaller for the stronger impurity potentials. 
This state has minimum and maximum along the $xz$-plane, the energy gap is small in the $yz$-plane and there is a weak maximum along the $\hat{y}$-axis in the $xy$-plane. This excitation energy spectrum is reflected in the thermal conductivity. As the $\kappa/T$ becomes finite, the in-plane anisotropy is very weak with slightly larger value along the $\hat{x}$-axis, as long as there is no node in  the $yz$-plane. As the impurity induced quasiparticles overcome the minima along the $\hat{y}$ axis, the thermal conductivity along the $\hat{y}$ axis starts to  dominate. In case of eight nodes, with four in the $xy$-plane and another four $yz$-plane, the thermal conductivity is always larger along the $\hat{y}$-axis. The $\hat{z}$-axis thermal conductivity remains  close to the $\hat{x}$-axis thermal conductivity.  For the $A_{1u}+irB_{3u}$ state, the thermal conductivity along the $\hat{x}$-axis and $\hat{y}$-axis show the same behavior as the $\kappa_{yy}/T$ and $\kappa_{xx}/T$ in the $A_{1u}+irB_{2u}$ state, respectively. 

For the $B_{1u}+irB_{3u}$ state, in the gapped phase \textit{i.e.} $r<1/\sqrt{2}$, the spectral gap on the Fermi surface is small in the $yz$-plane, and near the $\hat{y}$ axis in the $xy$-plane, which leads to larger $\kappa/T|_{T\rightarrow 0}$ along the $\hat{y}$ axis followed by the $\hat{z}$ direction, as shown in Fig. \ref{fig:k02}(i) and  Fig. \ref{fig:k02}(m), for $\tan \theta_s=2$ and $\tan \theta_s=5$, respectively. As a pair of quadruple nodes appears in the $yz$-plane, the relative anisotropy remains the same, as depicted in Fig.\ref{fig:k02}(j) and Fig.\ref{fig:k02}(n).  This state also shows an elusive quadratic node for $r=1$ along the $\hat{y}$-axis. Unlike the other states with linear point nodes, for this case the $\kappa/T|_{T\rightarrow 0}$ term remain finite along the nodes and for other directions residual thermal conductivity remains zero below the threshold disorder level. This state shows linear DOS at low energies, like superconductors with line nodes.  Finally, the nodes appear in the $xy$-plane for this state as $r$ goes beyond unity and remains closer to the $\hat{y}$-axis, and thermal conductivity along the $\hat{y}$-axis becomes dominant, while other two directions  show very similar $\kappa T_c/\kappa_N T$. As $r$ increases, the in-plane anisotropy decreases and the $\hat{x}$-axis $\kappa$ increases and becomes stronger along the $\hat{y}$ direction in $r\gg1$ limit. For the $B_{1u}+irB_{2u}$ state, the in-plane anisotropy found for the $B_{1u}+irB_{3u}$ state gets interchanged.

\begin{figure}
\includegraphics[width=1\columnwidth]{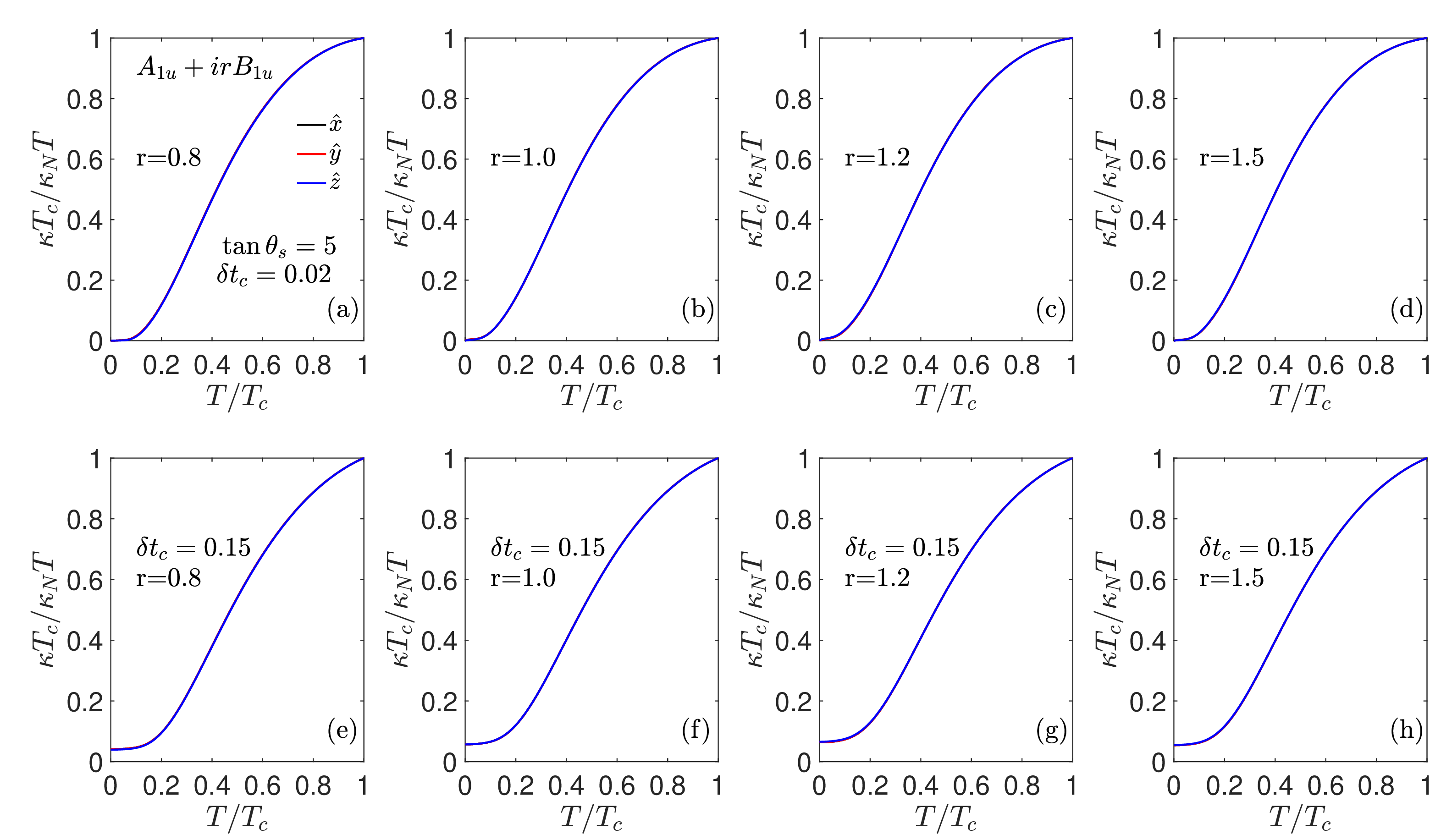} 
\caption{{\bf $A_{1u}+irB_{1u}$:}  Thermal conductivity normalized to its normal state value at $T_c$   for  the $A_{1u}+irB_{1u}$ shown as a function of temperature normalized to $T_c$ for scatterers with $\tan \theta_s =5$ for $\delta t_c=0.02$ from panels (a) to (d) and for $\delta t_c=0.15$ in panels (e) to (h) for various values of mixing parameter $r$.  } 
\label{fig:kte_1} 
\end{figure}
\subsection{Finite $T$ electronic thermal conductivity}
Now, we look at the temperature evolution of the normalized thermal conductivity for the nonunitary states. At very low temperatures, the elastic scattering by the impurities is the main mechanism of relaxation. However, as the temperature increases the inelastic scattering also becomes important, which we discuss in the subsequent section. Apart from the electronic contribution to thermal conductivity, phonon thermal conductivity can also become significant.    Here, we  focus on electronic thermal conductivity and the effect of impurity scattering on it, and the effect of underlying spectral nodes on the anisotropy in the thermal conductivity.   As shown in the Fig. \ref{fig:1u}, the thermal conductivity  shows a weak maximum as function of temperature at very low temperatures along the nodal directions. There is no evidence for such a feature in the experimental measurements\cite{Metz2019,suetsugu2023fully}; therefore, we have set $\theta_s = \tan^{-1}(5)$ for the rest of our discussion (see supplementary materials for the weaker impurity scatterers). We first look at the $A_{1u}+irB_{1u}$ state, which shows very isotropic residual thermal conductivity, and the thermal conductivity remains isotropic as a function of temperature, as shown in Fig. \ref{fig:kte_1}(a) to (d). For the clean system with $\delta t_c=0.02$, at very low temperatures, the thermal conductivity remains vanishingly small and as the temperature increases, $\kappa/T$ {increases $T^2$ at very low temperatures}. This behavior is observed for both nodal and gapped systems. One should keep in mind that in superconductors with point nodes, there are very few states available at the Fermi energy, as shown in Fig. \ref{fig:DOS1}. For dirtier systems such as $\delta t_c=0.15$, there are sufficient quasiparticle states at the Fermi level to give nonzero thermal conductivity, which is depicted in   Fig. \ref{fig:kte_1}(e) to (h). At low temperatures, $\kappa/T$ remains independent of temperature, and it begins to increase, once more quasiparticles become relevant for transport as the temperature increases. 
\begin{figure}
\includegraphics[width=1\columnwidth]{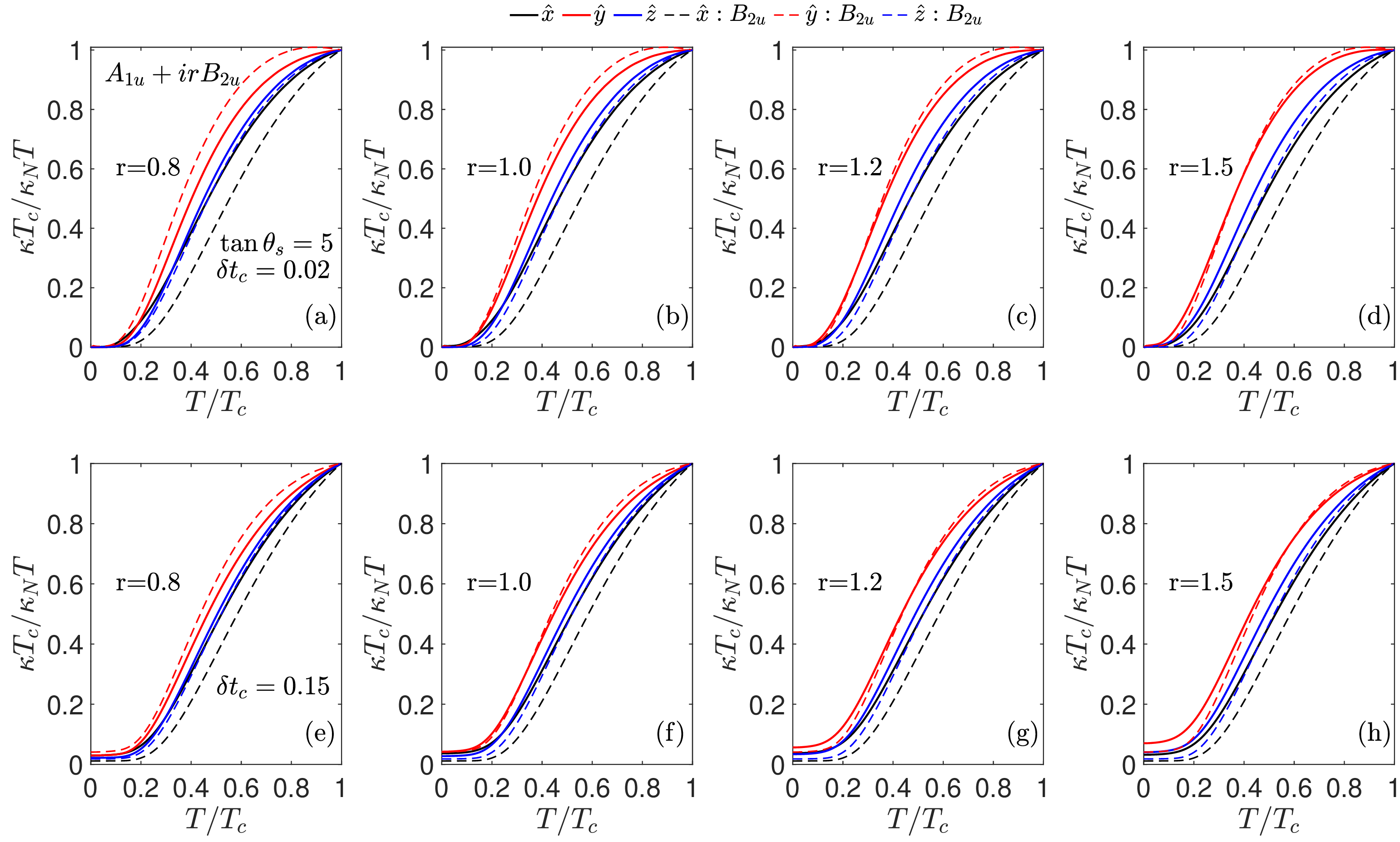} 
\caption{{\bf  $A_{1u}+irB_{2u}$:}  Thermal conductivity normalized to its normal state value at $T_c$   for  the $A_{1u}+irB_{2u}$ shown as a function of temperature normalized to $T_c$ for a scatterers with $\tan \theta_s =5$ for $\delta t_c=0.02$ from panels (a) to (d) and for $\delta t_c=0.15$ in panels (e) to (h) for various values of mixing parameter $r$. The dashed lines show the normalized thermal conductivities for the $B_{2u}$ state with same impurity parameters.  } 
\label{fig:kte_2} 
\end{figure}
\begin{figure}
\includegraphics[width=1\columnwidth]{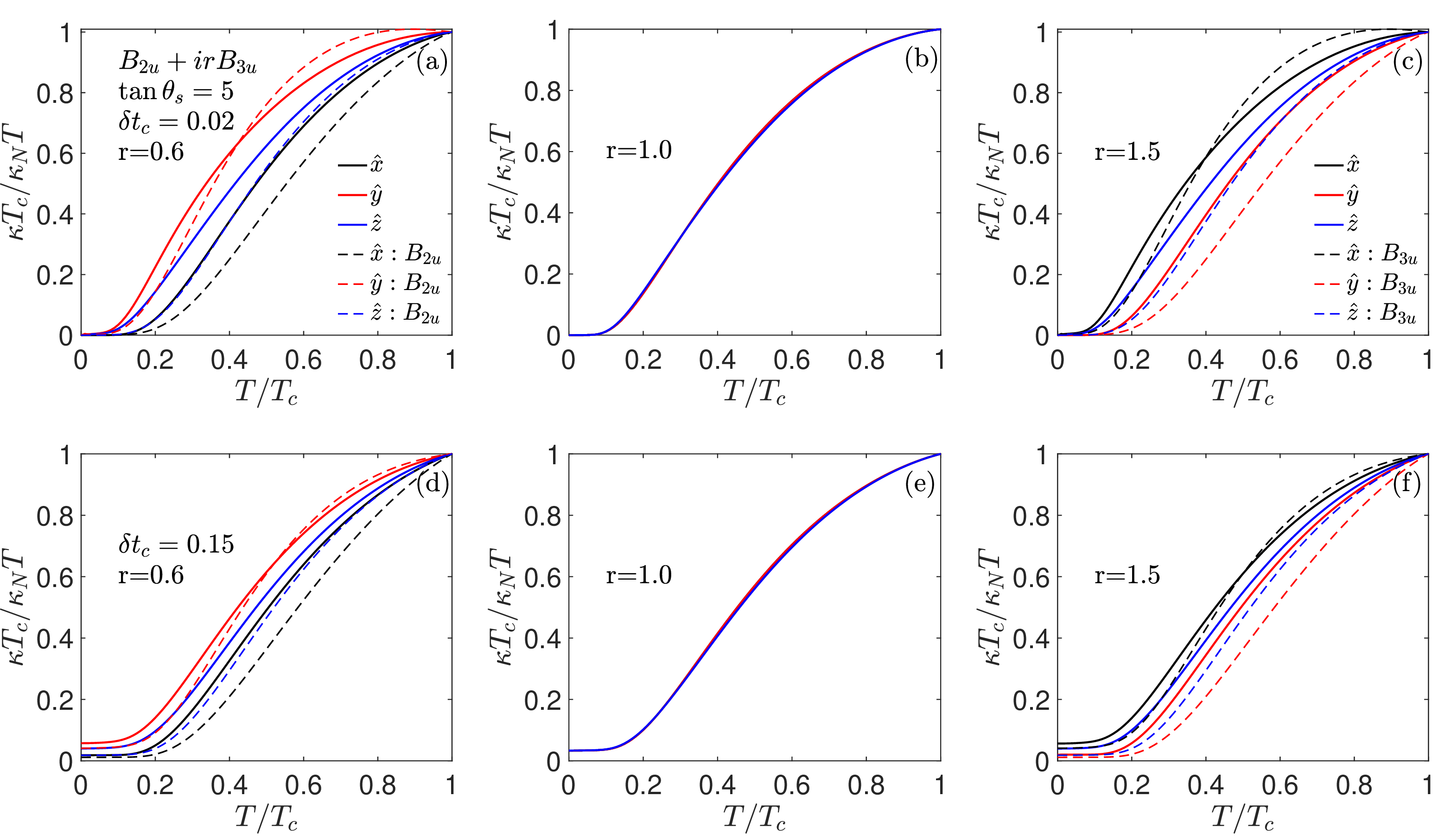} 
\caption{{\bf $B_{2u}+irB_{3u}$:}  Thermal conductivity normalized to its normal state value at $T_c$   for  the $B_{2u}+irB_{3u}$ shown as a function of temperature normalized to $T_c$ for a scatterers with $\tan \theta_s =5$ for $\delta t_c=0.02$ from panels (a) to (c) and for $\delta t_c=0.15$ in panels (d) to (f) for various values of mixing parameter $r$. The dashed lines show the normalized thermal conductivities for the $B_{2u}$ state in panels (a) and (d) and for the $B_{3u}$ state in panels (c) and (f). The impurity parameters are the same for the single component superconducting states.   } 
\label{fig:kte_3} 
\end{figure}

Next, we look at the $A_{1u}+irB_{2u}$ state, which shows enhanced thermal conductivity along the $\hat{y}$-axis. The zero temperature trends in anisotropy continue as the temperature increases, as shown in Fig. \ref{fig:kte_2}. For $\delta t_c=0.02$, $\kappa/T$ is close to zero and rises as temperature increases beyond a critical value. For the gapped case ($r<1$) and for the $\hat{x}$-axis nodes ($r=1$), $\kappa/T$ is slightly larger than other two directions, but as the temperature increases, the other two directions begin to increase and dominate because the temperature overcomes the gap minima in the $yz$-plane. For $r>1$, there are four point nodes which move towards the $\hat{y}$ axis, and it is reflected as a larger $\kappa/T$ along the $\hat{y}$-axis. For $r>\sqrt{2}$, four more nodes in $yz$-plane make $\kappa/T$ along $\hat{y}$ and $\hat{z}$ directions larger than the $\hat{x}$ direction.  For dirtier systems, thermal conductivity becomes finite and $T$-independent at very low temperature like the earlier case, but the anisotropy remains similar to its zero temperature limit. For $A_{1u}+iB_{3u}$, the behavior of $\hat{x}$ and $\hat{y}$ directions get interchanged (see supplementary materials). 

Next, we consider the $B_{2u}+irB_{3u}$ state, which is one of the FM nonunitary chiral states. For this state, there are four point nodes in $yz$-plane for $r< 1/\sqrt{2}$ and for $r> \sqrt{2}$, there is a quadruple pair of nodes in the $xz$-plane. Fig. \ref{fig:kte_3} shows the thermal conductivity as a function of temperature, and as expected, $\kappa/T$ along the nodal directions dominate. For $r<1$, the thermal conductivity is enhanced along the $\hat{y}$ and $\hat{z}$ directions, and as $r$ increases, the system becomes gapped with enhanced thermal conductivity in the $\hat{y}$ and $\hat{z}$ directions, but the anisotropy reduces. For $r=1$, this state shows a completely isotropic behavior and as $r$ goes beyond unity, this state show $\hat{x}$-axis dominated thermal response, as the nodes reappear in the $xz$-plane. The anisotropy remains qualitatively same as the temperature increases. 
\begin{figure}
\includegraphics[width=1\columnwidth]{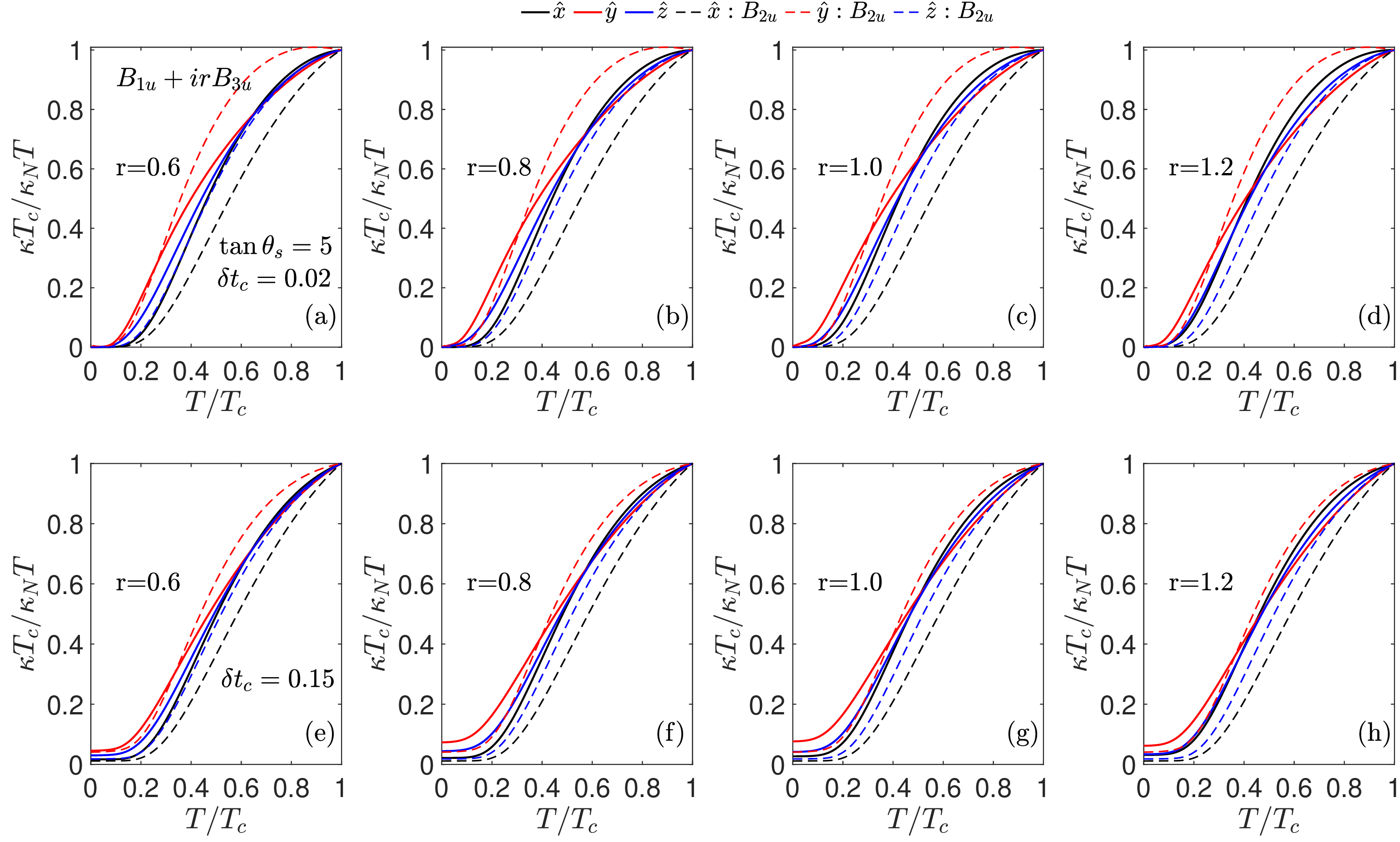} 
\caption{{\bf $B_{1u}+irB_{3u}$:}  Thermal conductivity normalized to its normal state value at $T_c$   for  the $B_{1u}+irB_{3u}$ shown as a function of temperature normalized to $T_c$ for a scatterers with $\tan \theta_s =5$ for $\delta t_c=0.02$ from panels (a) to (d) and for $\delta t_c=0.15$ in panels (e) to (h) for various values of mixing parameter $r$. The dashed lines show the normalized thermal conductivities for the $B_{2u}$ state with same impurity parameters.   } 
\label{fig:kte_4} 
\end{figure}
Finally, we look at the $B_{1u}+irB_{3u}$ state, which is another possible FM nonunitary state that shows enhanced thermal conductivity along the $\hat{y}$ directions. Here, the anisotropy changes significantly as the temperature increases. At low temperatures, $\hat{y}$ axis dominates due to its vicinity to the point nodes, however, as the temperature increases, we find enhancement of $\kappa T_c/\kappa_N T$ along the $\hat{x}$ directions. This happens because at the lower temperatures, the lower energy branch of the quasiparticle excitations $\sqrt{\xi_\k^2+\Delta_-^2}$  dominates, which has more quasi-particle states along the $\hat{y}$ directions, but at higher temperatures $\sqrt{\xi_\k^2+\Delta_+^2}$  branch of the quasiparticle excitations begin to contribute, which for this state has minima along the $\hat{x}$ axis. The overall anisotropy for this state is relatively less compare to other states, except those that show fully isotropic behavior as a function of temperature or disorder. For the $B_{1u}+irB_{2u}$ state, $\kappa/T$   along the $\hat{x}$ and $\hat{y}$ directions gets interchanged. 
\subsection{Inelastic scattering effects}
As we mentioned in the previous section, at very low temperatures \textit{i.e.} $T\ll T_c$, the elastic scattering from impurities is the only mechanism that determines the scattering rate. However, at higher temperature inelastic scattering from a bosonic mode is possible. We consider a simple scenario where there is a  dispersionless bosonic mod that couples with the fermions with an effective coupling constant $g_{fb}$. We further assume that the coupling does not depend on the fermion's spin degree of freedom. The lowest order self-energy for the fermions reads,
\begin{eqnarray}
\Sigma_{in}(i\omega_n, \mathbf{k}) &=& g^2_{fb} T \sum_{m,\mathbf{q}} G(i\omega_n-i\Omega_m , \mathbf{k}-\mathbf{q}) D(i\Omega_m,\mathbf{q}).
\label{Eq:Sf1_in}
\end{eqnarray}
Here $D$ is the bosonic Green's function, $\omega_n$ and $\Omega_m$ are the fermionic and bosonic Matsubara frequencies. After performing the Matsubara summation, we get,
\begin{eqnarray}
\Sigma_{in}(\omega)  &=&  -\frac{ g^2_{fb}}{2\pi} \int_{-\infty}^{\infty} dx \int_{-\infty}^{\infty} dy \frac{N(y) D''(x) }{x+y-\omega-i\eta} \left( \coth(\beta x/2) + \tanh(\beta y/2)\right).
\label{Eq:Sf2_in}
\end{eqnarray}
Here the real part of the self-energy contributes to mass renormalization and the imaginary part modifies the scattering rate, which is a function of quasiparticle energy and temperature. Noting that the \ute2 has very high effective mass, we ignore the real part of the self energy. The imaginary part of the self-energy is,
\begin{equation}
\Sigma_{in}''(\omega)  =  -\frac{ g^2_{fb}}{2} \int_{-\infty}^{\infty} dx N(\omega-x) D''(x) \left( \coth(\beta x/2) + \tanh(\beta (\omega-x)/2)\right),
\end{equation}
where $D''(x)\equiv x/(x^2+\Omega_0^2)$ is the bosonic DOS, and $\Omega_0$ is the characteristic energy scale associated  to the bosonic mode. In the context of \ute2, we expect $\Omega_0\gg T_c$, therefore, we approximate the bosonic DOS as $D''(x)\approx x/\Omega_0^2$. In the zero temperature limit, $\Sigma_{in}'' \propto \omega^{n+2}$, where the DOS for the superconducting state behaves like $\omega^n$ at low energies. For linear point nodes, $\Sigma_{in}'' \propto \omega^4$ and for quadratic point nodes or line nodes, it behaves like $\omega^3$.  Similarly, in the static limit ($\omega \rightarrow 0$), the imaginary part of the inelastic self-energy  reads,
\begin{equation}
\Sigma_{in}''(\omega=0,T)  =  -{ 2 g^2_{fb}} \int_{0}^{\infty} dx \frac{N(x) D''(x)}{\sinh \beta x}. 
\label{eq:Sf3_in}
\end{equation}
Since the thermal response integrand is peaked at $\omega=0$, we only retain the temperature dependence of the inelastic self-energy at $\omega=0$. At very low temperatures, the inelastic scattering rate behaves likes $T^4$ for the point nodes and $T^3$ for the line nodes in the static limit. This is sufficient to understand the qualitative effect of the inelastic scattering. The prefactor in Eq. \eqref{eq:Sf3_in} is fixed by the value of the inelastic scattering rate at $T_c$.
\begin{figure}
\includegraphics[width=1\columnwidth]{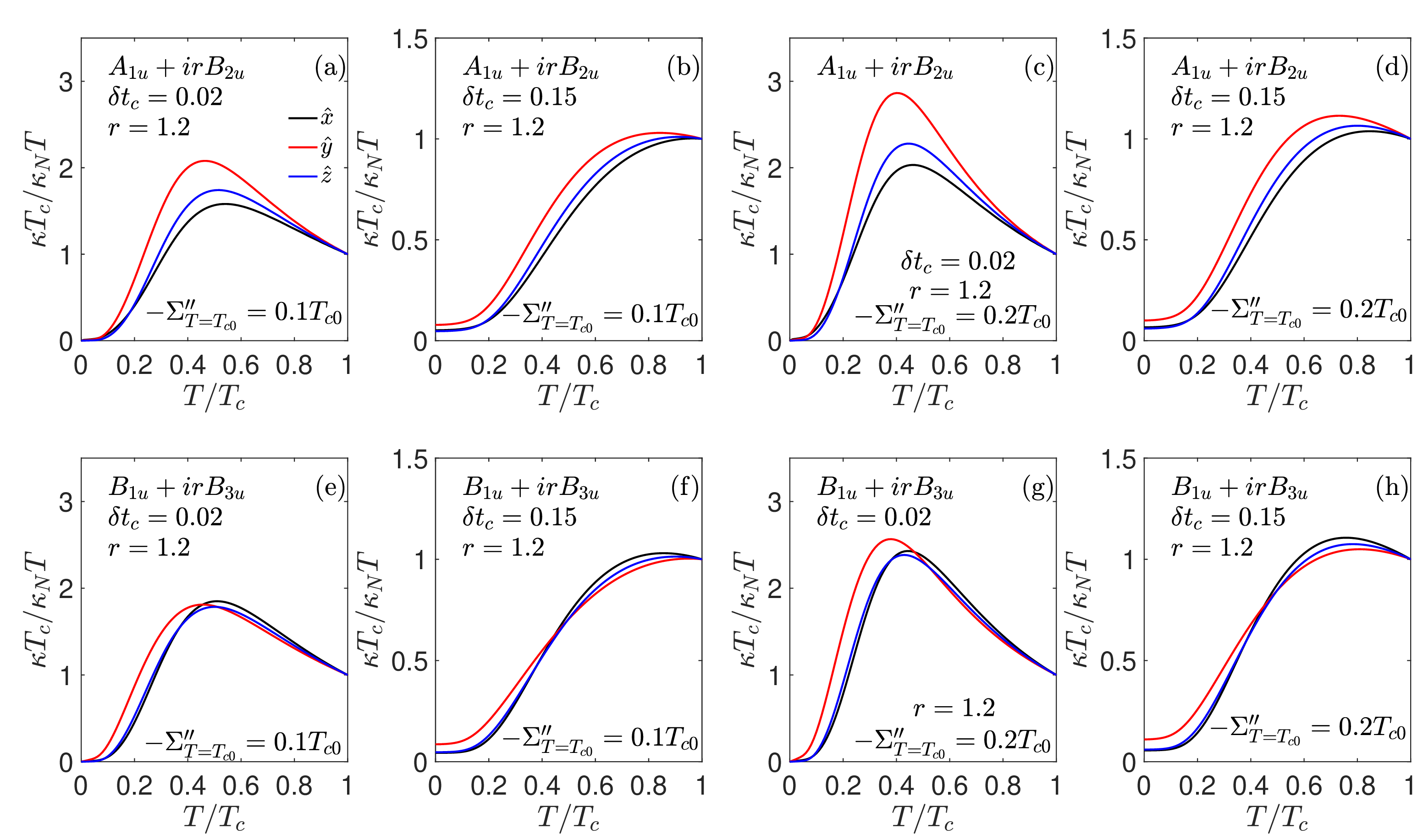} 
\caption{{\bf Effect of inelastic scattering scattering:} Panels (a) to (d) show temperature dependence of normalized thermal conductivity for the $A_{1u}+irB_{2u}$ state for $r=1.2$. Panels (e) to (h) show the temperature dependence of the normalized thermal conductivity for the $B_{1u}+ir B_{3u}$ state. The imaginary part of the the inelastic self energy at $T=T_{c0}$ and relative reduction in $T_c$ are indicated in each panel. For brevity, we denote $\Sigma^{\prime\prime}_{in}(T=T_{c0})$ as $\Sigma^{\prime\prime}_{T=T_{c0}}$. The $s$-wave scattering phase is $\tan^{-1}(5)$.} 
\label{fig:k_in} 
\end{figure}

Fig. \ref{fig:k_in} shows the effect of the inelastic scattering on the electronic thermal conductivity. We show the results for two cases, as the rest of the cases are qualitatively similar. The first row of the Fig. \ref{fig:k_in} shows the normalized thermal conductivity for the $A_{1u}+irB_{2u}$ state, which show enhanced thermal conductivity along the $\hat{y}$-axis. Note, the normal state thermal conductivity includes the inelastic scattering. We have fixed the value of $r$ at $1.2$, which leads to four point nodes in the $xy$-plane. The primary effect of inelastic scattering is formation of a peak below $T_c$. This peak  appears first  for the directions that have larger thermal conductivity, as shown in Fig. \ref{fig:k_in}(a) for $\delta =0.02$. As the system becomes dirtier, the peaks get smeared as depicted in Fig. \ref{fig:k_in}(b) for $\delta t_c=0.15$.  If the strength of inelastic scattering is increased, the peaks also strengthen, as shown in  Fig. \ref{fig:k_in}(c) and it may survive in dirtier systems, as illustrated in Fig. \ref{fig:k_in}(d).
The similar trends continue for the other states also. In panels \ref{fig:k_in}(e) to (h), we show the normalized thermal conductivity for the $B_{1u}+irB_{3u}$ state, which shows qualitatively same behavior as a function of disorder and inelastic scattering strength. However, this state shows a nonmonotonic variation of anisotropy, and the peaks for different directions remain quite close to each other. 
\section{Summary \& Concluding Remarks}
We studied low energy quasiparticle excitations and the thermal transport for the single component and two components pairing states allowed by the irreducible representations of  the \d2h point group symmetry, which is relevant for the orthorhombic \ute2 crystals. Since \d2h has only one dimensional representations,  the four single component  pairing states corresponding to the irreducible representations  can not break the time reversal symmetry, and describe unitary triplet pairing states. Therefore, we also considered the pairing states that  are  combinations of two of the four irreducible representations using a single mixing parameter $r$.  We examined  all  six two component superconducting states  as  a function of $r$ on a cylindrical Fermi surface, which describe either gapped states or state with spectral point nodes depending on the value of the mixing parameter. The spectral point nodes are not necessarily the zeros of order parameters, but these are the points on the Fermi surface hosting quasiparticle excitations. Of course, the spectral nodes are identical to the gap nodes in the case of single component or unitary states. These six states can be divided into AF or FM categories depending on the Fermi surface average  of the Cooper pair spin-moment, which vanishes for the AF states and remains finite for the FM states. Except for the $A_{1u}+irB_{1u}$ state, all other states are chiral on the cylindrical Fermi surface, as the average angular momentum of the Cooper pairs remains finite. 

After introducing the one parameter model for the two component states, we calculated the effect of impurity scattering within the self-consistent t-matrix approximation. One of the new findings is the spin-dependent impurity scattering rate for the chiral states. This happens due to finite quasiparticle spin densities for the chiral states. This can be interpreted as accumulation of magnetization near the impurity sites and this leads to qualitative changes in the quasiparticle excitation spectrum. For the two component states, the nodes are accidental not symmetry imposed like the single component states, and the spin-dependent self-energy or the impurity pinned magnetization can change the position of the spectral nodes. In principle, the removal of spectral point nodes  by impurity scattering is possible in the chiral superconducting states, but we have not found such effect for the cases that we consider. Next, we calculate the thermal conductivity using the Kubo formula for the thermal current response function. Due to spin-dependent impurity self-energies, the thermal conductivity  significantly differs from the thermal conductivity reported for the unitary states. We examine the thermal transport for  all the single and double component states that are possible for the \d2h point group. 

We have considered a single band with cylindrical Fermi surface in our theoretical calculations. The quantum oscillation experiment reports two cylindrical Fermi surfaces, where one of them is an electron-like and the other one is a hole-like Fermi surface with comparable effective masses\citep{Aoki2022_QO}. However, we expect our analysis to be valid for two band system too, because the interband scattering is always pair-breaking due to odd-parity order parameters. The impurity will renormalize the quasiparticle energies, there will not be any off-diagonal impurity self-energy. Therefore, multiple bands will lead to higher impurity  scattering rate, this should not affect the anisotropy of the thermal transport. There are also some speculations about a closed Fermi surface near the $\mathbf{Z}$ point, therefore, we also considered a spherical Fermi surfaces (see appendix). The key qualitative difference is the possibility of nodes along the $\hat{z}$ axis and strong thermal conductivity along that directions, however, there is no experimental data available to support that scenario. 

Based on our thermal transport study and some recent experimental data, we can identify some states that could possibly describe the gap structure  in \ute2. Definitive conclusions are not possible at this time due to a lack of sufficient direction-dependent data on the newer samples, but we can make some qualitative statements and rule out some states. For the $A_{1u}+irB_{1u}$ state on a cylindrical Fermi surface, the normalized thermal conductivity shows isotropic behavior as a function of impurity scattering in the zero temperature limit and in its temperature dependence for a fixed disorder level. The limited data that are available for the thermal conductivity indicates weak in-plane anisotropy, but not absolute isotropic behavior\cite{Metz2019}. The thermal conductivity measurements by Suetsugu \textit{et al.}\cite{suetsugu2023fully} claims a fully gapped superconducting state.  {In contrast, another independent thermal conductivity measurements by Hayes \textit{et al.} find evidence for point nodes without finding any residual thermal conductivity in the zero temperature limit\cite{Hayes2024}.} Therefore, absence of finite zero temperature limit thermal conductivity in high quality samples is not sufficient to rule out point nodes.  Other probes such as the field dependence of specific heat suggest a superconducting state with nodes closer to $\hat{y}$-axis\cite{lee2023anisotropic}. The superfluid density measurement indicate stronger low energy quasiparticle excitations along the $\hat{y}$ directions and weakest along the $\hat{x}$ axis\cite{Ishihara2023}. The relative anisotropy in the penetration depth measurements is weaker compare to the single component states. We find that the $A_{1u}+irB_{2u}$ state, the $B_{2u}+irB_{3u}$  state with dominant $B_{2u}$  component and the $B_{1u}+irB_{3u}$ state  show stronger quasi-particle excitations along the $\hat{y}$-axis. The  $B_{1u}+irB_{3u}$ state  shows change of anisotropy as a function of temperature, which can be used to distinguish it from other two states. This state also shows a quadratic point node for $r=1$, which shows linear DOS at low energies, hence, it can be also ruled out. The two ferromagnetic states show higher thermal conductivity along the $\hat{y}$ direction than the $B_{2u}$ state, which has point nodes along the $\hat{y}$-axis.

As we mentioned earlier that the phonon thermal conductivity could be significant, especially in the low $T_c$ samples. At very low temperatures, the scattering of phonons from defects dictates the phonon mean free path, hence, the phonon thermal conductivity\cite{Klemens1955,BRT1959,Uher1990}. Phonon thermal conductivity is expected to be insignificant in the high quality samples due to low concentration of defects. Therefore,  a systematic  measurement of thermal conductivity along all three directions in the samples with high residual resistivity ratios  is highly desirable.  One common feature among all these superconducting phases is zero $\kappa/T$ in the zero temperature limit in clean samples. For sufficient disorder, even point nodal states acquire a very small residual $kappa/T$, which would require very careful low-T measurements to detect.

\section*{Acknowledgments}
The authors are  grateful to S. Anlage, I. Hayes, T. Metz,  J. P. Paglione, and T. Shibauchi  for useful discussions.  P.J.H. was supported by the NSF under DMR-2231821.

\setcounter{figure}{0}
\renewcommand{\thefigure}{A\arabic{figure}}%
\appendix

\section{Thermal Conductivity}

The heat current vertex $\check{\mathbf{j}}_{Q_{i}}$ along the i$^{th}$ ($i=\hat{x},\hat{y},\hat{z}$) reads,
\begin{eqnarray}
\check{\mathbf{j}}_{Q_{i}} &=& (\omega + \Omega/2 ) \left[ v_{Fi} \tau_3 \otimes \sigma_0   \right],
\label{Eq:heat_current}
\end{eqnarray}
where $i$ denotes the direction of the heat current in the real space along $\hat{x}/\hat{y}/\hat{z}$. The thermal conductivity $\kappa$ is obtained from the current-current correlation function for the thermal current\citep{AmbegaokarGriffin:1994}. After doing the  Matsubara summation, thermal conductivity becomes,
\begin{equation}
\frac{\kappa}{T} = \frac{1}{\pi}\int_{-\infty}^{\infty} \frac{d\omega}{T} \frac{\omega^2}{T^2} \left(- \frac{d n_{F} (\omega) }{d\omega}\right) \frac{\mathrm{Tr}\left[ \check{\mathcal{A}}(\k,\omega) \check{\mathbf{j}}_{Q_{i}} \check{\mathcal{A}}(\k,\omega)  \check{\mathbf{j}}_{Q_{i}}\right]}{2}.
\label{Eq:kappa0}
\end{equation}
Here a factor of half accounts for the double counting of the spins while taking the trace, $n_F$ is the Fermi function and $\check{\mathcal{A}}$ is the spectral function, which reads,
\begin{eqnarray}
\check{\mathcal{A}}&=& \frac{  \left[ \check{\mathbf{G}}(\k,\omega+i\eta)-\check{\mathbf{G}}(\k,\omega-i\eta) \right] }{2i}= \frac{  (\check{\mathbf{G}}^R - \check{\mathbf{G}}^A)}{2i},
\end{eqnarray}
where the impurity dressed Green's function  $\check{\mathbf{G}}$ is,
\begin{eqnarray}
\check{\mathbf{G}}&=&\begin{pmatrix}
\hat{\mathbf{G}}_{11} & \hat{\mathbf{G}}_{12} \\
\hat{\mathbf{G}}_{21} & \hat{\mathbf{G}}_{22}
\end{pmatrix},
\end{eqnarray}
where the block Green's functions in the spin-space are,
\begin{eqnarray}
\hat{\mathbf{G}}_{11} &=&  \frac{L_0 + \mathbf{L}_1 \cdot \bsig}{\widetilde{\mathbf{D}}}, \label{G11_dr} \\
\hat{\mathbf{G}}_{22}&=& \frac{L_0(\xi \rightarrow -\xi) + \mathbf{L}_1(\xi \rightarrow -\xi) \cdot \bsig^T}{\widetilde{\mathbf{D}}}, \label{G22_dr} \\
 \hat{\mathbf{G}}_{12} &=& \left[ 2\xi (\bSig \cdot \dv ) + (\rw^2 - \xi^2 - \Delta_0^2 |\dv |^2 + \bSig \cdot \bSig ) \dv \cdot \bsig  + i \Delta_0^2 (\q \times \dv )  \cdot \bsig - 2 (\bSig \cdot \dv ) \bSig \cdot \bsig \right. \nonumber \\
 & & \left. + 2 i \rw (\bSig \times \dv)\cdot \bsig \right] 
 \frac{i\sigma_y \Delta_0}{\widetilde{\mathbf{D}}},  \label{G12_dr} \\
 \hat{\mathbf{G}}_{21} &=&  \left[ -2 \xi  \bSig \cdot \dv^\ast  +  ( \rw^2 -\xi^2- \Delta_0^2 | \dv |^2  + \bSig\cdot\bSig  ) \dv^\ast \cdot \bsig^T    - i \Delta_0^2 (\q \times \dv^\ast)  \cdot \bsig^T -2 \bSig \cdot \dv^\ast \bSig \cdot \bsig^T \right. \nonumber \\
 & & \left.  -2i\rw (\bSig \times \dv^\ast)\cdot \bsig^T \right]  \frac{i \sigma_y \Delta_0}{ \widetilde{\mathbf{D}}},  \label{G21_dr}
\end{eqnarray}
where 
\begin{eqnarray}
L_0 &=& X_0 a_0 - \Delta_0^2 b_0 |\dv |^2 - \Delta_0^2 \q \cdot \bSig, \label{Eq:L0} \\
\mathbf{L}_1 &=& (X_0+ \Delta_0^2 | \dv |^2 ) \bSig + \Delta_0^2 b_0 \q - \Delta_0^2 \bSig \cdot \dv \dv^\ast  - \Delta_0^2 \bSig \cdot \dv^\ast \dv,  \label{Eq:L1} 
\end{eqnarray}
Here $a_0=\rw- \xi$,  $b_0=\rw+ \xi$, $X_0 = b_0^2-\bSig\cdot \bSig$ and the denominator $\widetilde{\mathbf{D}} = (\xi^2 + \rQ_+^2) ( \xi^2 + \rQ_-^2)$, where $\rQ_\pm$ is,
\begin{eqnarray}
\rQ_\pm^2 &=& \Delta_0^2 | \dv |^2 -\rw^2 - \bSig \cdot \bSig  \pm \sqrt{\Delta_0^4 \q \cdot \q + 4\rw^2 \bSig \cdot \bSig + 4\Delta_0^2 \rw \q \cdot \bSig -  4 \Delta_0^2   (\bSig \cdot \dv) (\bSig \cdot \dv^\ast) }
\end{eqnarray}
 First we  evaluate the trace term,
\begin{eqnarray}
\frac{1}{2}\mathrm{Tr}\left[\check{\mathcal{A}}(\k,\omega) \check{\mathbf{j}}_{Q_{i}} \check{\mathcal{A}}(\k,\omega)  \check{\mathbf{j}}_{Q_{i}}  \right] &=& v_{Fi}^2 \frac{1}{2}\mathrm{Tr} [T_1 ], \nonumber \\
T_1 &=& \hat{\mathcal{A}}_{11}\hat{\mathcal{A}}_{11}- \hat{\mathcal{A}}_{12} \hat{\mathcal{A}}_{21}- \hat{\mathcal{A}}_{21} \hat{\mathcal{A}}_{12}+ \hat{\mathcal{A}}_{22} \hat{\mathcal{A}}_{22}.
\label{Eq:tr1}
\end{eqnarray}
Note, the Green's function have complex gap function, but the poles of the Green's function depend on the sign of $\tilde{\omega}$, which is renormalized due to disorder. We can express $\check{\mathbf{G}}''$ as $(\check{\mathbf{G}}-\check{\mathbf{G}}^*)/2i$, and only a combination of $\check{\mathbf{G}}$ and $\check{\mathbf{G}}^*$ will have poles on the opposite side of the real axis to make a non-zero contribution to the integral over $\xi$.
For $\xi$ integration, it is useful to express the numerator and the denominator in terms of polynomials. The numerator can be written as,
\begin{eqnarray}
g_4(\xi) &=& b_0 \xi^6 + b_1 \xi^4 + b_2 \xi^2 + b_3, \label{Eq:NumT1} \\
b_0&=&1,\\
b_1 &=&  (| \rw|^2 -\Delta_0^2 |\dv |^2 + | \bSig |^2 ) + \mathrm{Re}\left[ \tilde{Q}_+^2 + \tilde{Q}^{2}_- \right] \\
b_2&=&\frac{1}{4} | \tilde{Q}_+^2 + \tilde{Q}^{2}_- |^2 + (|\rw |^2 + |\bSig |^2 - \Delta_0^2 |\dv |^2) \mathrm{Re}\left[ \tilde{Q}_+^2 + \tilde{Q}^{2}_- \right] +  3\Delta_0^4 \q \cdot \q + 4 |\bSig |^2 ( |\rw |^2 +\mathrm{Re}[\rw^2] )  \nonumber \\
& & - 4\Delta_0^2 \left( | \bSig \cdot \dv |^2 + |\bSig \cdot \dv^\ast |^2 + \mathrm{Re}[(\bSig \cdot \dv) (\bSig \cdot \dv^\ast) ] \right) \nonumber \\
& &+ 4 |\rw |^2 \mathrm{Re}[ \bSig \cdot \bSig] + 4\Delta_0^2 \mathrm{Re}[ \rw \q \cdot \bSig^\ast  + 2 \rw \q \cdot \bSig] \\
b_3 &=&  ( |\rw |^2 -\Delta_0^2 |\dv |^2 ) \left[ |\alpha_+ |^2 + \Delta_0^4 \q \cdot \q \right]- 2 \Delta_0^4 \q \cdot \q \mathrm{Re}[\alpha_+] + \mathcal{Y}(\bSig) \\
 {\mathcal{Y}}(\bSig)  &=&  \left\lbrace |(\rw^2 + \Delta_0^2 |\dv |^2 - \bSig \cdot \bSig)  |^2 -\Delta_0^4 \q \cdot \q   \right\rbrace |\bSig |^2 - 2\Delta_0^2 \mathrm{Re}[ (\alpha_- - \alpha_-^\ast ) \rw \q\cdot \bSig^\ast ] \nonumber \\
 & & - 4\Delta_0^2 (|\rw |^2+ |\bSig |^2 -\Delta_0^2 |\dv |^2 )  \mathrm{Re}[ \rw \q \cdot \bSig^\ast ] +  2 \Delta_0^4 \mathrm{Re}\left[| \q \cdot \bSig |^2 -(\q \cdot \bSig )^2+ \bSig \cdot \bSig \q \cdot \q \right] \nonumber \\
 && + 2\Delta_0^2\mathrm{Re} \left[(  | \rw |^2- \rw^2 ) - ( |\bSig |^2 -\bSig \cdot \bSig ) \right] ( |\bSig \cdot \dv |^2 + |\bSig \cdot \dv^\ast |^2 ) -4 \Delta_0^2 |\dv |^2 |\rw |^2 | \bSig |^2  \nonumber \\
 & &  +4\Delta_0^2 \mathrm{Re}\left[ \alpha_+ (\bSig^\ast \cdot \dv^\ast ) (\bSig^\ast \cdot \dv ) \right] - 4\Delta_0^2 \mathrm{Re}[ (\rw^2 - \Delta_0^2 |\dv |^2 ) \bSig^\ast \cdot \bSig^\ast ],
\end{eqnarray}
where $\alpha_\pm = \rw^2 -\Delta_0^2 |\dv |^2 \pm \bSig \cdot \bSig$, and the denominator can be expressed as,
\begin{eqnarray}
|\mathbf{D}(\xi)|^2 &=& h_4(\xi) h_4(-\xi), \\
h_4(\xi) &=& (\xi-i Q_+)  (\xi- i Q_+^*)   (\xi- i Q_-)   (\xi- i Q_-^*) \\
&=& a_0 \xi^4 + a_1 \xi^3 + a_2 \xi^2 +a_3 \xi +a_4, \\
a_0 &=& 1, \\
a_1 &=& -i (Q_+ + Q_+^* + Q_- + Q_-^* )= -2i\mathrm{Re}(Q_+ + Q_-) = i c_1 , \\
a_2 
&=& -|Q_+|^2 - |Q_-|^2 - 4 \mathrm{Re}(Q_+)  \mathrm{Re}(Q_-) = - c_2\\
a_3 
&=& 2i |Q_+|^2  \mathrm{Re}(Q_-) +  2i |Q_-|^2  \mathrm{Re}(Q_+) = -i c_3, \\
a_4 &=& Q_+ Q_+^* Q_- Q_-^* = |Q_+|^2 |Q_-|^2 = c_4.
\end{eqnarray}
Here $h_4(\xi)$ is defined to ensure poles on the upper half of the plane for the retarded Green's functions. Using\citep[Eq. 3.112.5]{GradshteynRyzhik}
\begin{eqnarray}
\int_{-\infty}^{\infty} dx \frac{g_4(x)}{h_4 (x) h_4 (-x)} &=& i\pi \frac{b_0 (-a_1 a_4 + a_2 a_3 )- a_0 a_3 b_1 + a_0 a_1 b_2 + a_0 b_3 (a_0 a_3 -a_1 a_2 )/a_4}{a_0 (a_0 a_3^2 + a_1^2 a_4 - a_1 a_2 a_3)}, \\
&=&  i\pi \frac{ (-a_1 a_4 + a_2 a_3 )-  a_3 b_1 + a_1 b_2 +  b_3 ( a_3 -a_1 a_2 )/a_4}{ ( a_3^2 + a_1^2 a_4 - a_1 a_2 a_3)},  \nonumber \\
&=& \pi \frac{ (-c_1 c_4 + c_2 c_3 )+ c_3 b_1 + c_1 b_2 +  b_3 ( -c_3 +c_1 c_2 )/c_4}{ ( c_3^2 + c_1^2 c_4 - c_1 c_2 c_3)}
\end{eqnarray}

Finally, we have the thermal conductivity,
\begin{equation}
\frac{\kappa_{ii}}{T} =  \int_{-\infty}^{\infty} d\omega \frac{\omega^2}{T^2} \left(- \frac{d n_{F} (\omega) }{d\omega}\right) \left\langle  N_0 v_{Fi}^2 \frac{ (-c_1 c_4 + c_2 c_3 )+ c_3 b_1 + c_1 b_2 +  b_3 ( -c_3 +c_1 c_2 )/c_4}{ ( c_3^2 + c_1^2 c_4 - c_1 c_2 c_3)} \right\rangle_{FS}
\label{Eq:kappa_f}
\end{equation}
\section{Gap equation and transition temperature}
We consider a simple separable pairing potential 
\begin{eqnarray}
\mathbf{V}_{\alpha \beta ; \alpha' \beta'} (\k,\k') = -V (\mathbf{d}(\k)\cdot \boldsymbol{\sigma} i\sigma_y)_{\alpha \beta}  (\mathbf{d}(\k')\cdot \boldsymbol{\sigma} i\sigma_y)^\dagger_{\alpha' \beta'}
\end{eqnarray}
here $\alpha,~\beta,~\alpha',~\beta'$ denote spin and $V(>0)$ is the strength of pairing. This pairing potential gives a single transition temperature for the mixed two component order parameters. In principle, it is possible to construct an effective pairing potential to get more than one transition, however, such scenario is excluded in the current discussion.
The gap equation can be written as,
\begin{eqnarray}
\Delta_0 &=& \pi \lambda_{pair} T \sum_{\omega_n} \left[ \frac{1}{2}\left( \frac{1}{\tilde{\mathtt{Q}}_+}+\frac{1}{\tilde{\mathtt{Q}}_-}\right) |\dv |^2 
 - \frac{\Delta_0^2 \q \cdot \q - 2 \bSig\cdot \dv \bSig \cdot \dv^\ast + 2i\rw_n \q \cdot \bSig}{\tilde{\mathtt{Q}}_+ \tilde{\mathtt{Q}}_- \left( \tilde{\mathtt{Q}}_+ + \tilde{\mathtt{Q}}_-\right) } \right]\Delta_0
\end{eqnarray}
where $ \lambda_{pair}=N_0 V$ and $\tilde{\mathtt{Q}}_\pm$ is,
\begin{eqnarray}
\tilde{\mathtt{Q}}_\pm^2 &=& \Delta_0^2 | \dv |^2 +\rw^2_n - \bSig \cdot \bSig \nonumber \\
& & \pm \sqrt{\Delta_0^4 \q \cdot \q - 4\rw^2_n \bSig \cdot \bSig + 4i \Delta_0^2 \rw_n \q \cdot \bSig -  4 \Delta_0^2   (\bSig \cdot \dv) (\bSig \cdot \dv^\ast)}
\end{eqnarray}
We solve the gap equation using Matsubara Green's function. In Matsubara Frequencies, $\bSig \propto i \Delta_0^2$, therefore it will drop out near $T_c$ and the linearized  gap equation becomes,
\begin{equation}
\Delta_0 = \lambda (2\pi T) \sum_{\omega_n>0}^{\Omega_c} {\Delta_0}\left\langle |\mathbf{d}|^2 \left( \frac{1}{\omega_n+ \Gamma_N} \right)  \right\rangle_{FS},
\end{equation}
where $\Gamma_N$ is the normal state scattering rate. Using $\theta \equiv \langle |\mathbf{d}|^2  \rangle_{FS} $, we can express $T_{c0}$,
\begin{equation}
T_{c0} = 1.13 \Omega_c \exp \left( -\frac{1}{\lambda \theta}\right),
\end{equation}
and the critical value of $\Gamma_N$ to completely kill the superconductivity as,
\begin{eqnarray}
\Gamma_{cric} &=& \frac{1}{1.13}T_{c0} = 0.885 T_{c0},
\end{eqnarray}
which is same as the critical value of impurity scattering rate  that kills $d$-wave superconductivity. The general equation to determine $T_{c}$ is,
\begin{eqnarray}
\ln \left( \frac{T_c}{T_{c0}}\right) &=& \Psi\left[ \frac{1}{2}\right] - \Psi\left[ \frac{1}{2} + \frac{\Gamma_N}{2\pi T_{c}}\right].
\end{eqnarray}
Here $\Psi$ denotes the digamma function and we have assumed $T_{c0},\Gamma_N \ll \Omega_c$. In the weak disorder limit ($\Gamma_N \ll T_{c0}$),
\begin{equation}
T_{c} \approx T_{c0} \left[ 1- \frac{\pi}{4} \frac{\Gamma_N }{T_{c0}}+ \frac{7\zeta(3)}{4\pi^2} \left( \frac{\Gamma_N }{T_{c0}} \right)^2\right]
\end{equation}
\section{Some additional results}
\subsection{Density of states: Nonunitary states on cylindrical Fermi surface}
\begin{figure}[htbp]
\begin{center}
\includegraphics[width=0.49\linewidth]{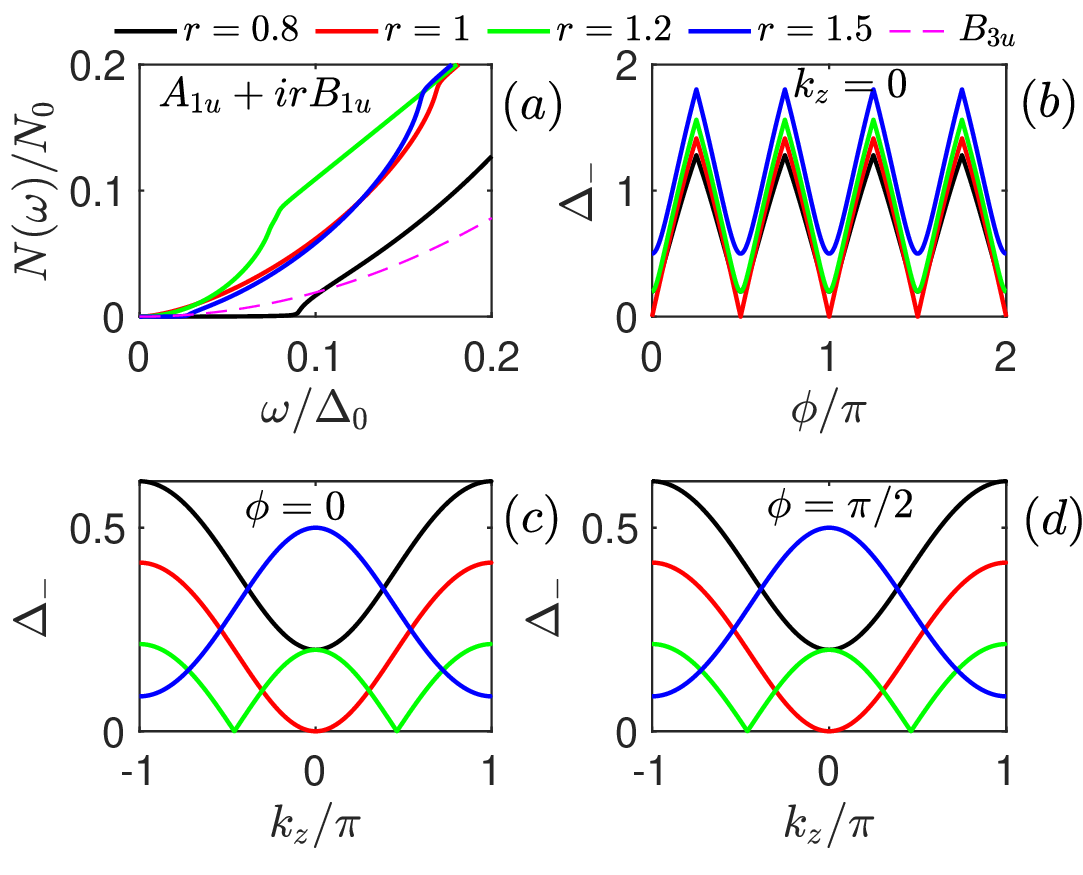}
\includegraphics[width=0.49\linewidth]{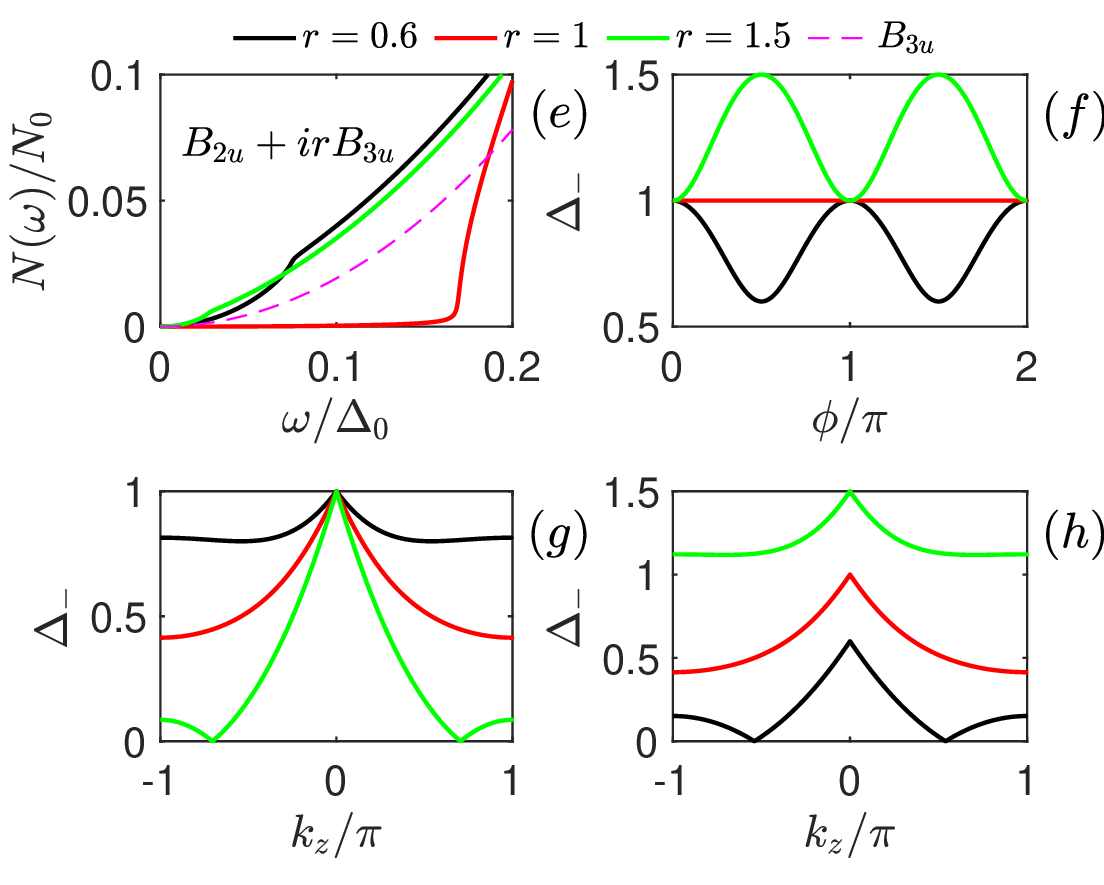}
\end{center}
\caption{ The DOS for the  $A_{1u}+irB_{1u}$ state shown in panel (a) with corresponding $\Delta_-$ structure's in the $xz$ and $yz$ planes in panels (b) and (c), respectively. The DOS for  the  $B_{2u}+irB_{3u}$ state  is shown in panel(d) alongside the $\Delta_-$  in the $xy$-plane in panel(e) and in the $yz$-plane in panel(f).  }\label{fig:s1}
\end{figure}

Fig. \ref{fig:s1} shows the low energy density of state (DOS) for the $A_{1u}+irB_{1u}$ state and the $B_{2u}+irB_{3u}$  state for various values of the mixing parameter $r$. The former is an antiferromagnetic (AF) nonunitary state, while the latter is a ferromagnetic nonunitary state. For both these states, the possibility of $\hat{z}$-axis nodes is prevented by the topology of the Fermi surface, which is assumed to be open along the $\hat{z}$-axis. 
\begin{figure}
\includegraphics[width=1\columnwidth]{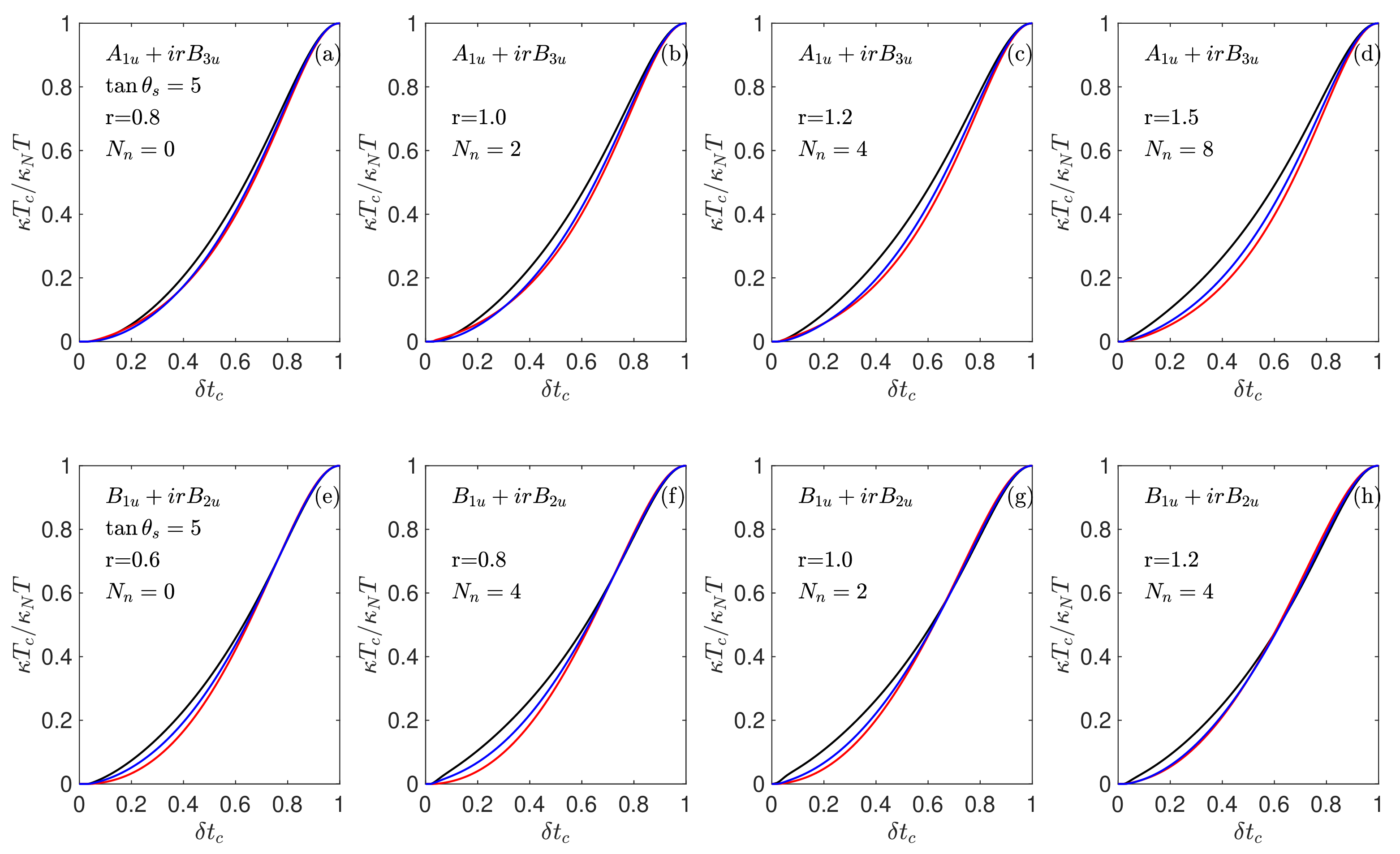} 
\caption{ Zero temperature limit $\kappa T_c/\kappa_N T$ for the $A_{1u}+irB_{3u}$ in panels (a) to (d) and for the $B_{1u}+irB_{2u}$    state in panels (e) to (h) for various values of mixing parameter $r$. The $s$-wave scattering phase shift is $\theta_s =\tan^{-1}(5)$ for all the panels.   } 
\label{fig:kt0_s2} 
\end{figure}
\subsection{Thermal conductivity for the $A_{1u}+irB_{3u}$ and $B_{1u}+irB_{2u}$  states}
\begin{figure}
\includegraphics[width=1\linewidth]{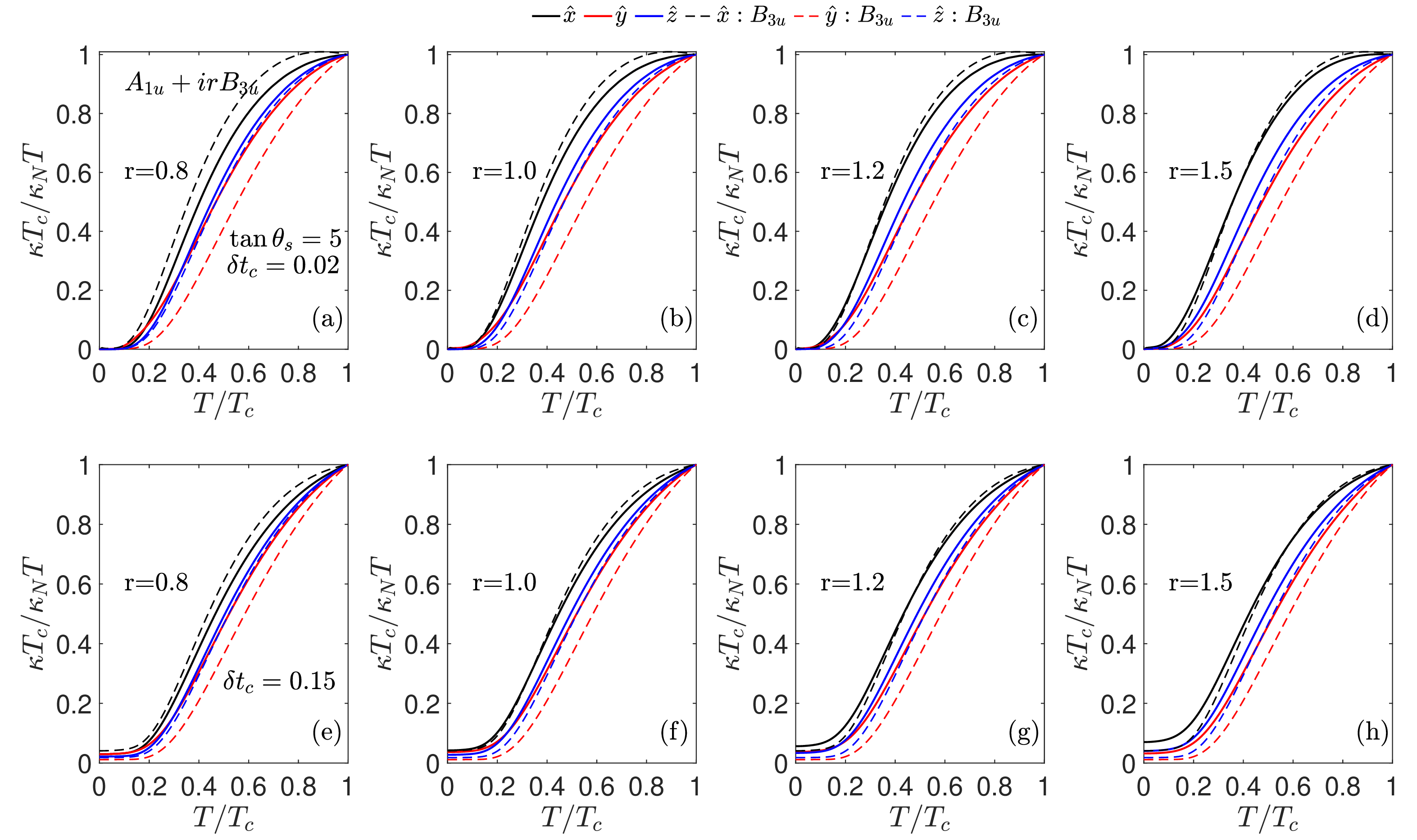} 
\includegraphics[width=1\linewidth]{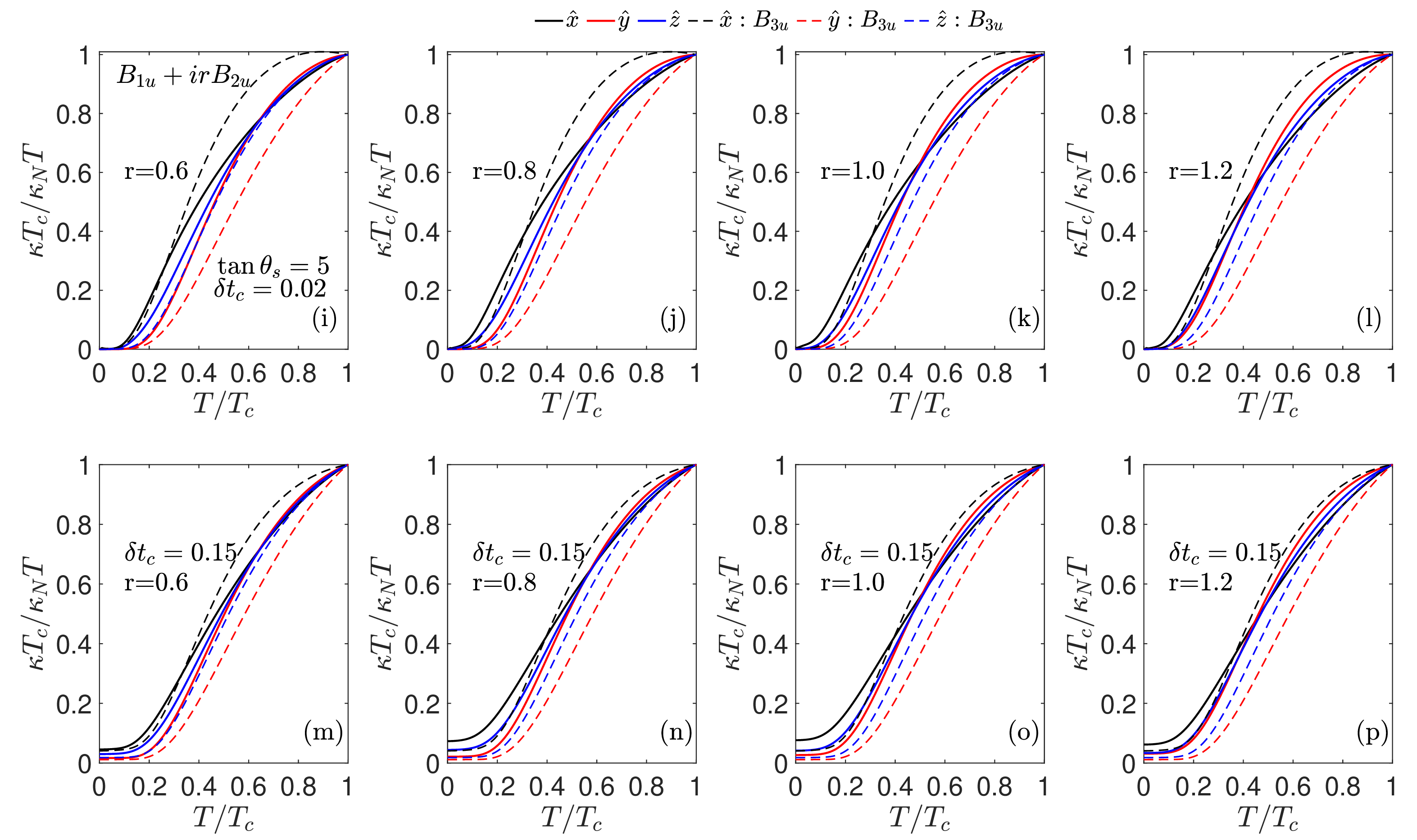} 
\caption{ Thermal conductivity normalized to its normal state value at $T_c$   for  the $A_{1u}+irB_{3u}$ shown as a function of temperature normalized to $T_c$ for a scatterers  for $\delta t_c=0.02$ from panels (a) to (d) and for $\delta t_c=0.15$ in panels (e) to (h) for various values of mixing parameter $r$. The normalized thermal conductivity for  the $B_{1u}+irB_{2u}$ shown as a function of temperature normalized to $T_c$ for $\delta t_c=0.02$ from panels (i) to (l) and for $\delta t_c=0.15$ in panels (m) to (p). The $s$-wave scattering phase shift is $\tan ^{-1}(5)$. } 
\label{fig:kte_s3} 
\end{figure}

Fig. \ref{fig:kt0_s2} shows the zero temperature limit thermal conductivity as a function of relative $T_c$ reduction $\delta t_c (\equiv 1-T_c/T_{c0})$ for the $A_{1u}+irB_{3u}$ in panels (a) to (d)  and for the $B_{1u}+irB_{2u}$  states  in panels (e) to (h). The $s$-wave scattering phase shift is $\tan ^{-1}(5)$. The temperature evolution for these states in shown in Fig. \ref{fig:kte_s3}.
\begin{figure}
\includegraphics[width=1\columnwidth]{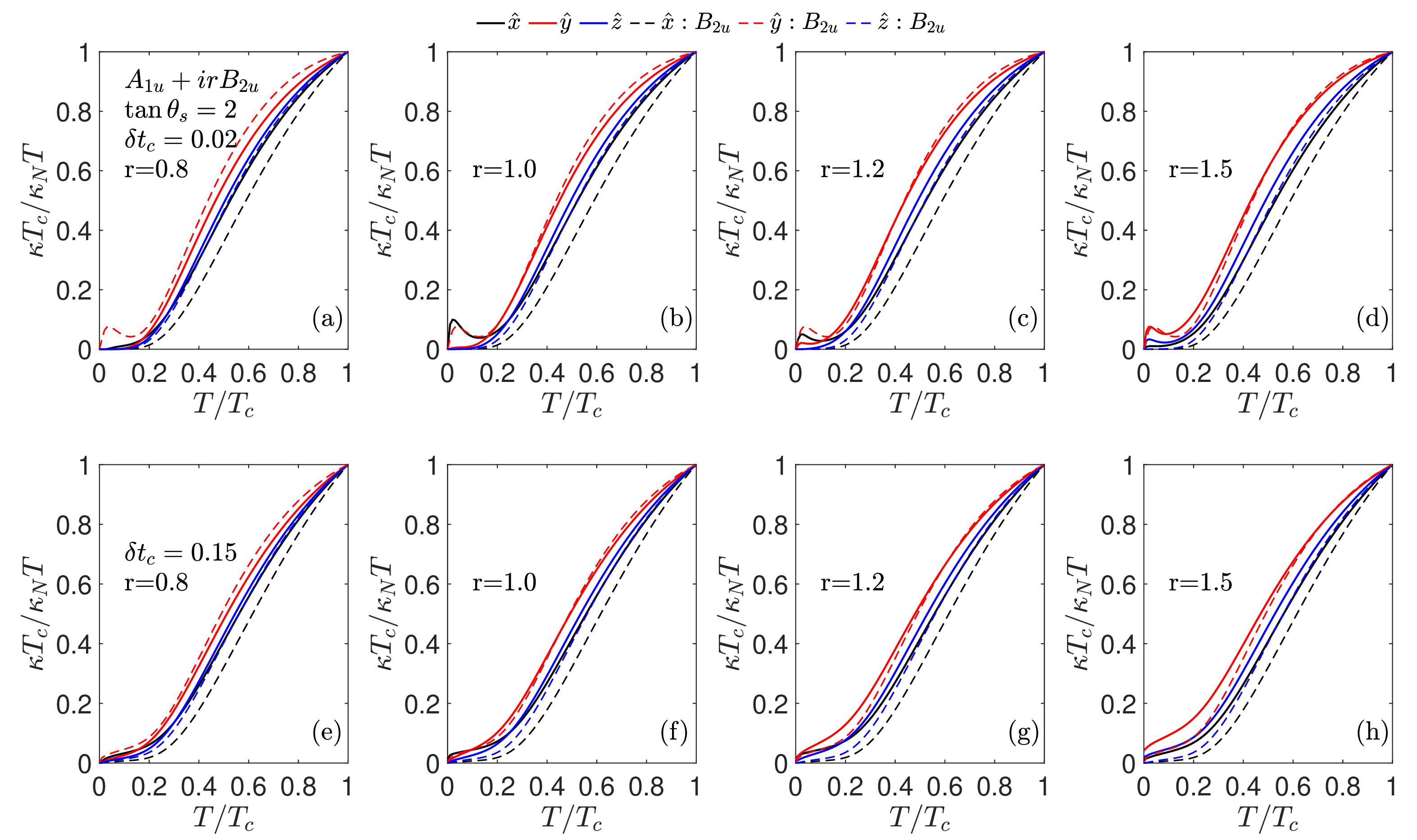} 
\includegraphics[width=1\columnwidth]{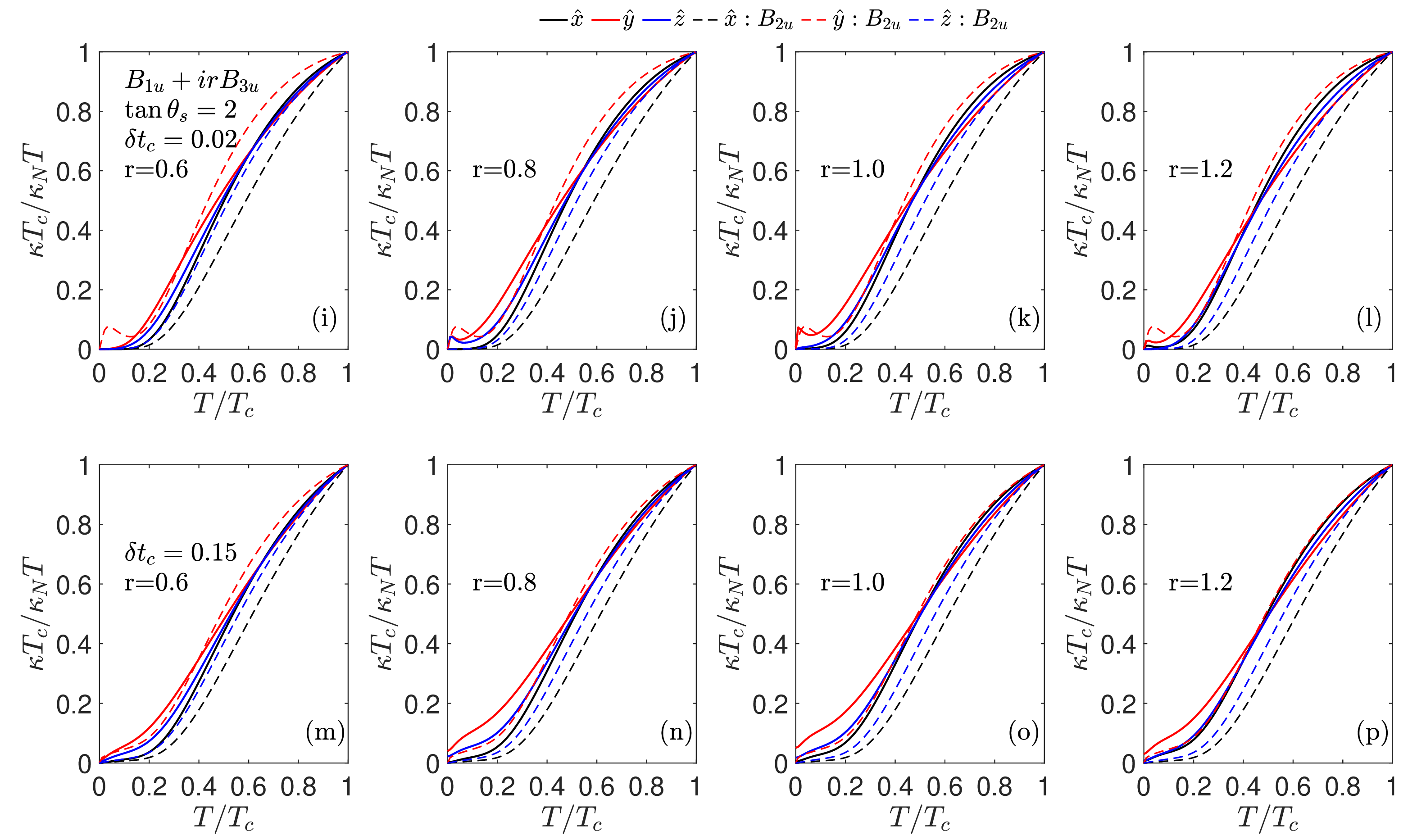} 
\caption{Thermal conductivity normalized to its normal state value at $T_c$   for  the $A_{1u}+irB_{2u}$ shown as a function of temperature normalized to $T_c$ for a scatterers  for $\delta t_c=0.02$ from panels (a) to (d) and for $\delta t_c=0.15$ in panels (e) to (h) for various values of mixing parameter $r$. The normalized thermal conductivity for  the $B_{1u}+irB_{3u}$ shown as a function of temperature normalized to $T_c$ for $\delta t_c=0.02$ from panels (i) to (l) and for $\delta t_c=0.15$ in panels (m) to (p). The $s$-wave scattering phase shift is $\tan ^{-1}(2)$.  } 
\label{fig:kte_s1} 
\end{figure}
\subsection{Effect of weak sactterers on thermal conductivity}
Fig. \ref{fig:kte_s1}  shows the temperature dependence of thermal conductivity normalized to its normal state value for the $A_{1u}+irB_{2u}$  state and for the $B_{1u}+irB_{3u}$ state. The $s$-wave scattering phase shift is $\tan ^{-1}(2)$. For each state, thermal conductivity is calculated for $\delta t_c=0.02$ and $\delta t_c=0.15$.
\begin{figure*}[b]
\includegraphics[width=0.49\columnwidth]{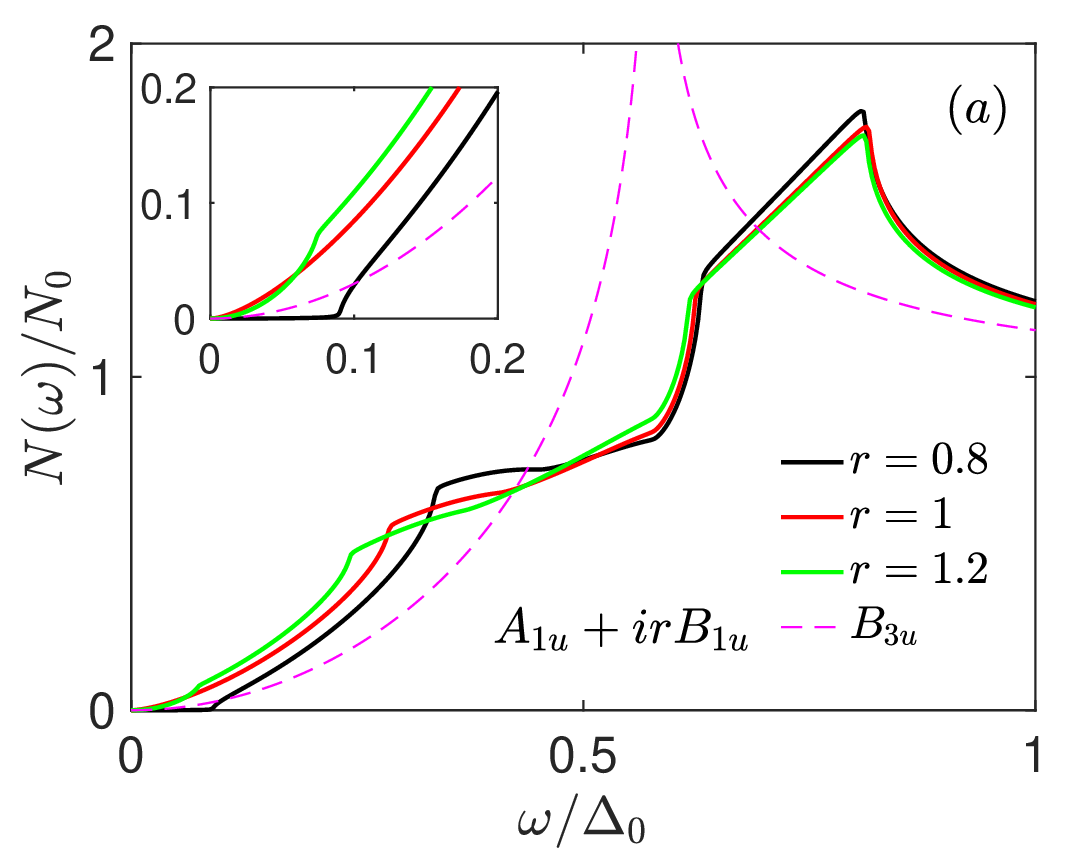} 
\includegraphics[width=0.49\columnwidth]{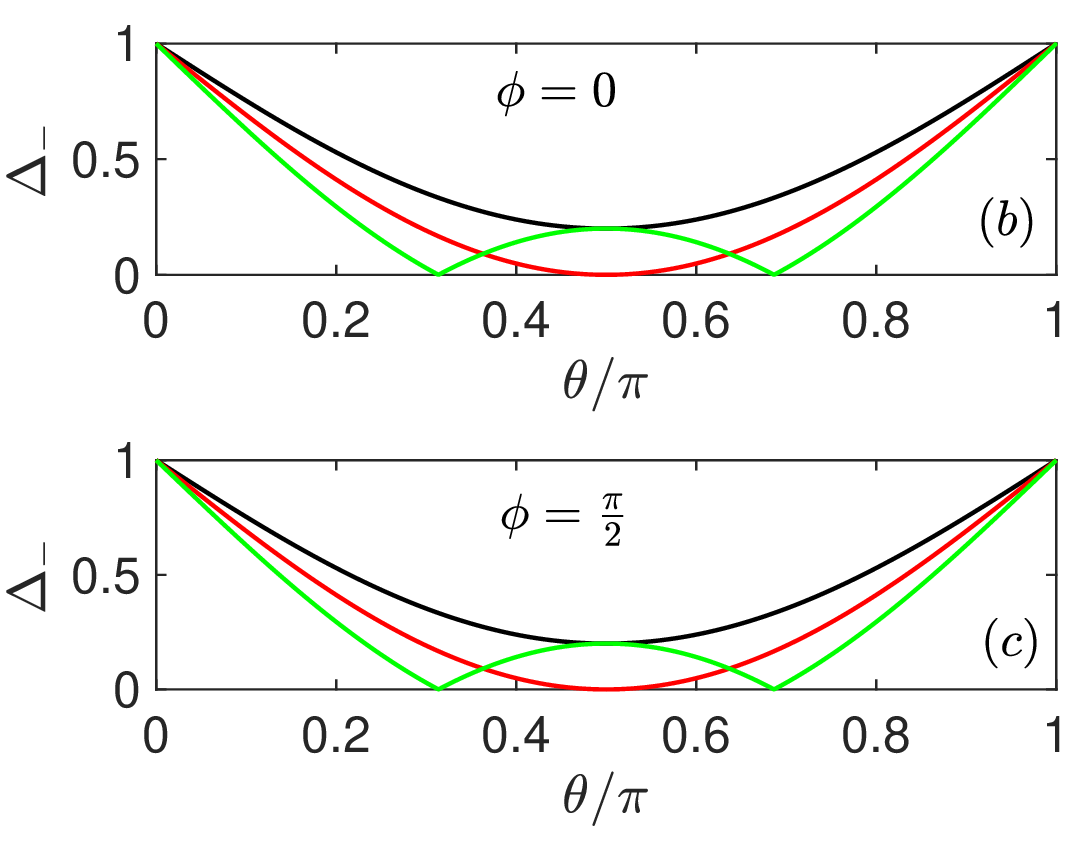} 
\includegraphics[width=0.49\columnwidth]{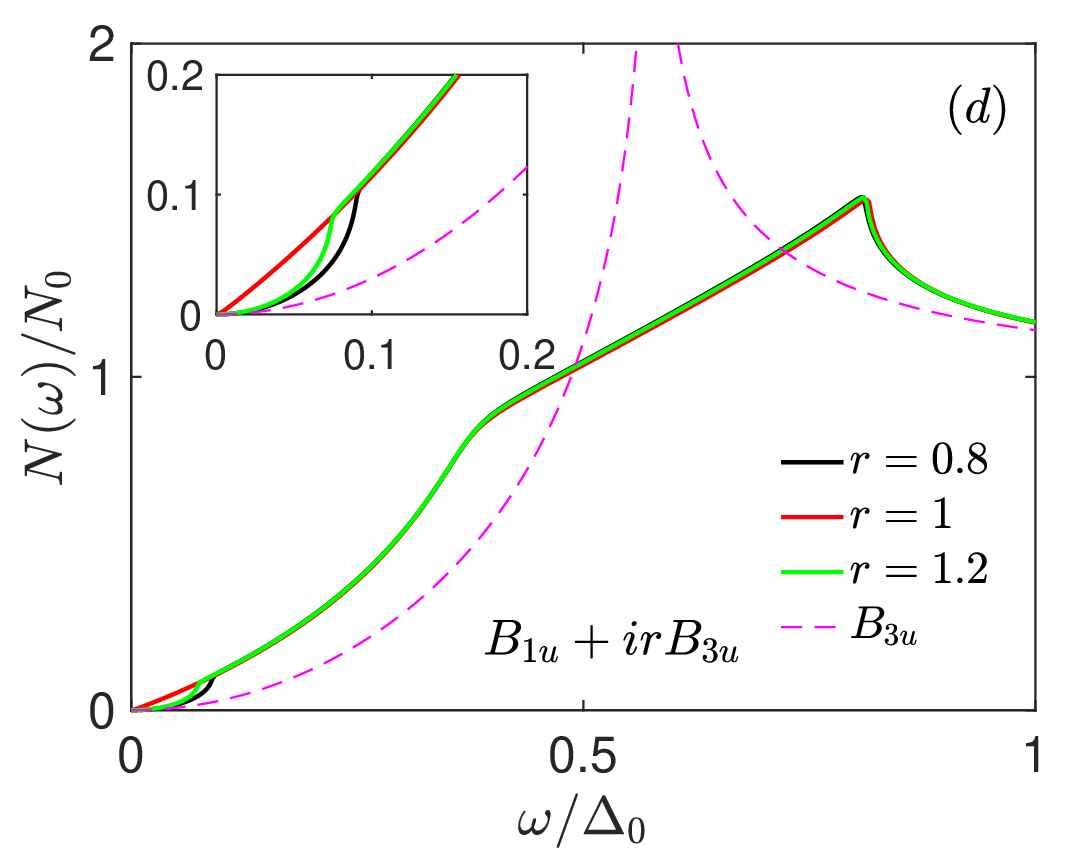} 
\includegraphics[width=0.49\columnwidth]{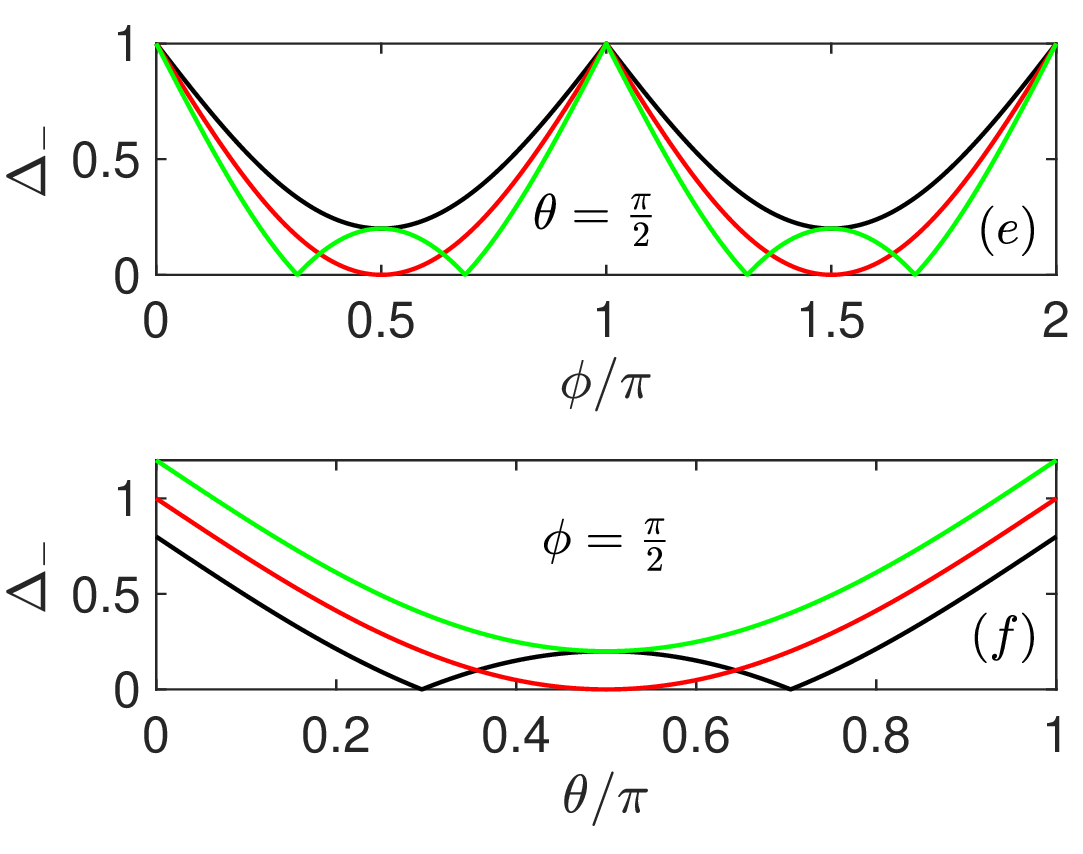} 
\caption{Panel (a) shows the DOS per spin for the $A_{1u}+irB_{1u}$ state for different values of the mixing parameter $r$. The inset shows the energy dependence of DOS at lower energies. The dashed line illustrates  the DOS for unitary $B_{3u}$ state has been shown for comparison.  Panel (b) and (c) show the variation of $\Delta_-$ for different values of $r$ in the $xy$ and $yz$ planes, respectively. Panel (d) shows the DOS per spin for the  $B_{1u}+irB_{3u}$ state, and the gap structures  in the  $xy$ and $xz$ planes are shown in panels (e) and (f). } 
\label{fig:DOS2} 
\end{figure*}
\section{Spherical Fermi surface}
\begin{figure}
\includegraphics[width=1\columnwidth]{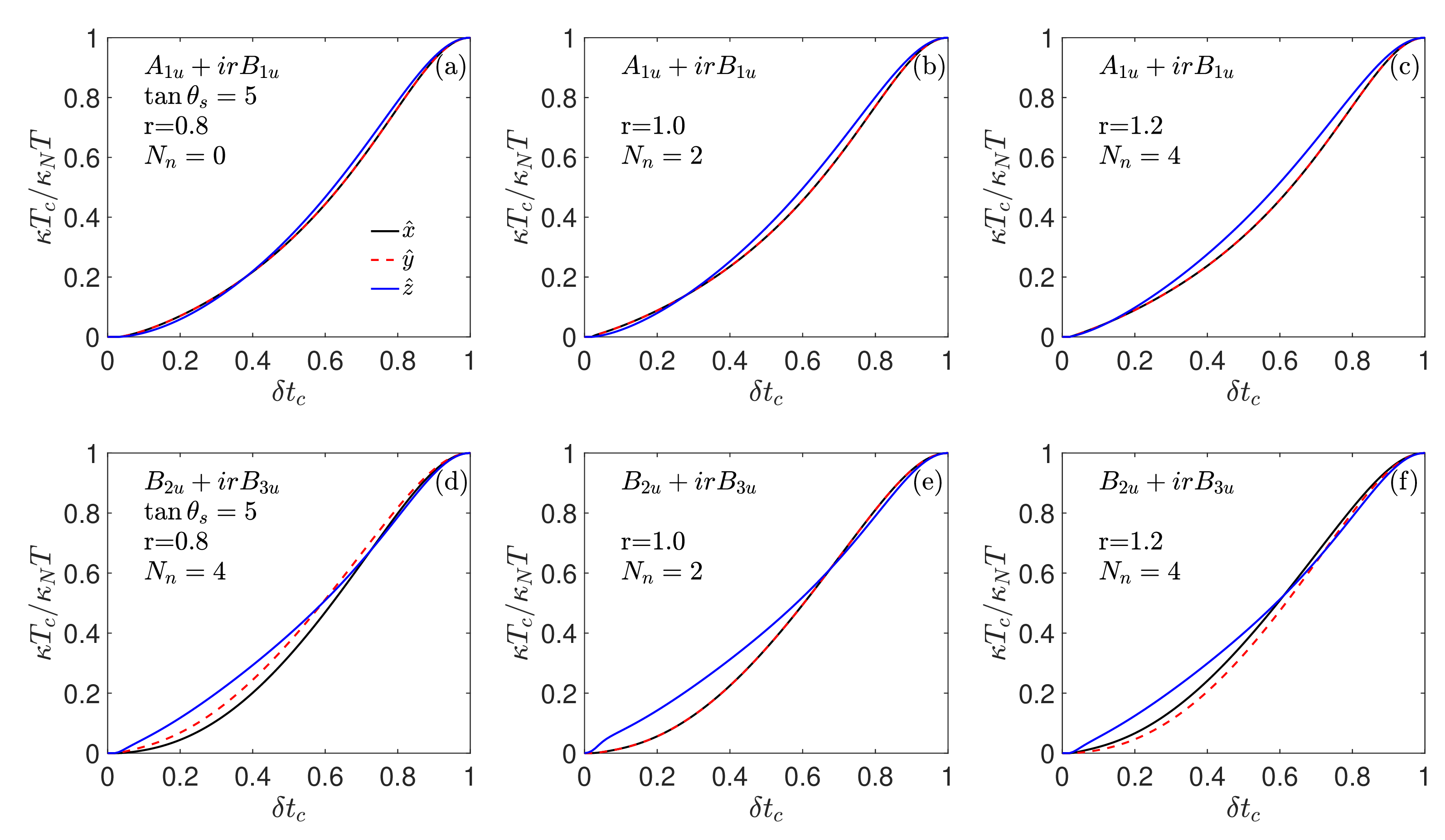} 
\caption{ Zero temperature limit $\kappa T_c/\kappa_N T$ for the $A_{1u}+irB_{1u}$ state in panels (a) to (c) and for the $B_{2u}+irB_{3u}$    state in panels (d) to (f) for various values of mixing parameter $r$. The Fermi surface is assumed to be a three-dimensional sphere. The number of nodes $N_n$ is indicated for each case.  } 
\label{fig:k0_sph} 
\end{figure}
\begin{figure}[t]
\includegraphics[width=1\columnwidth]{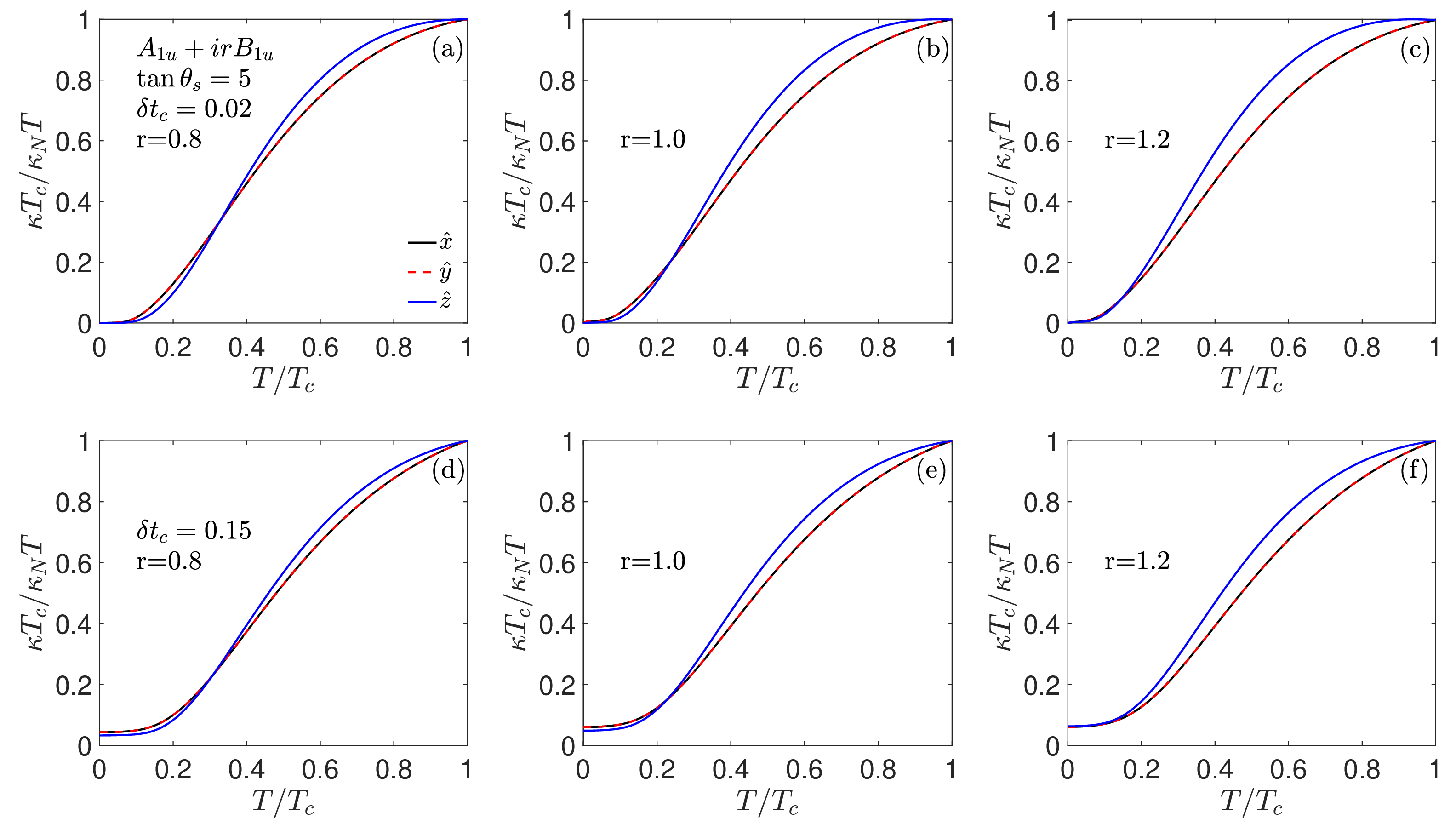} 
\includegraphics[width=1\columnwidth]{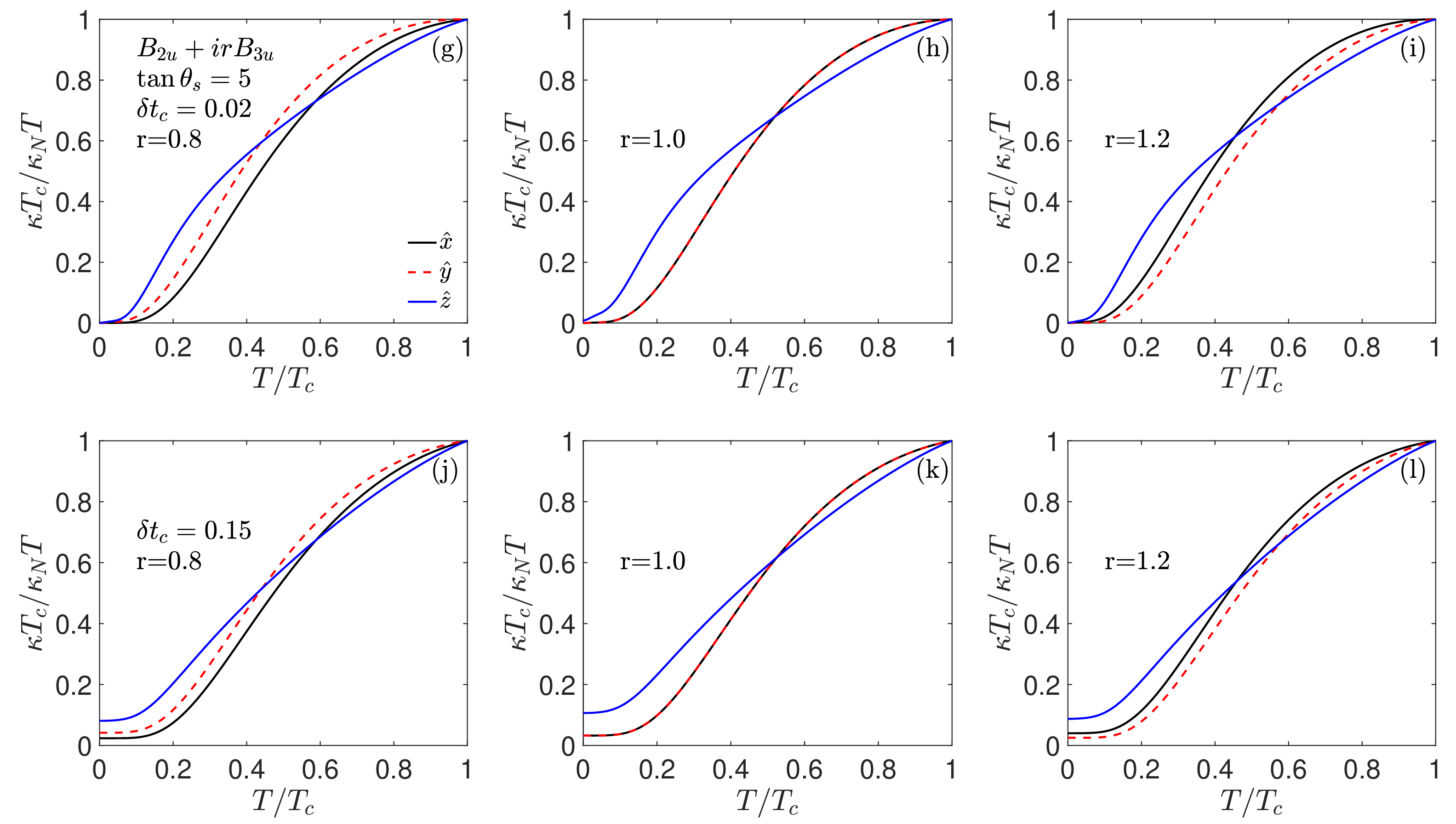} 
\caption{ Temperature dependence of  $\kappa T_c/\kappa_N T$ for the $A_{1u}+irB_{1u}$ state in panels (a) to (c) for $\delta t_c =0.02$ and for for $\delta t_c =0.15$ in panels (d) to (f) for various values of mixing parameter $r$. The Fermi surface is assumed to be a three-dimensional sphere and $\tan \theta=5$. Temperature dependence of  $\kappa T_c/\kappa_N T$ for the $B_{2u}+irB_{3u}$ state in panels (g) to (i) for $\delta t_c =0.02$ and for $\delta t_c =0.15$ in panels (j) to (l) for various values of mixing parameter $r$. } 
\label{fig:kt_sph} 
\end{figure}
Now we consider the two component states on a spherical Fermi surface.  Fig. \ref{fig:DOS2}(a) shows the DOS for the  $A_{1u}+irB_{1u}$ state, which is an AF and non-chiral state. The variation of $\Delta_-$ for this state is shown in the Fig. \ref{fig:DOS2} (b) and (c).  In case of a spherical Fermi surface, all AF states are non-chiral and all the FM nonunitary states are chiral. For $r<1$ , the $A_{1u}+irB_{1u}$ state remains gapped with deep minima along the $\hat{x}$ and $\hat{y}$ directions in the $k_z=0$ plane, and for $r=1$, four nodes appear along the $\hat{x}$ and $\hat{y}$ directions and show quadratic density of states. For $r>1$, nodes split into eight along the $\pm\hat{z}$ directions, where the polar angle of nodal position becomes $\tan \theta = \pm\sqrt{r^2-1}$. For large values of $r$, nodes move towards the $\hat{z}$ axis. The DOS remains quadratic at low energies. The DOS remains the same for other AF nonunitary states, but the location of minima and nodes changes. 

Next, we show the DOS of the FM nonunitary state $B_{1u}+irB_{3u}$ in Fig. \ref{fig:DOS2}(d) with $\Delta_-$ for this state in panel (e) and (f). For $r<1$, there are four point nodes in the $xz$ plane at $\cot \theta =\pm r/\sqrt{1-r^2}$ and for $r>1$, there are four nodes in the $xy$ plane at $\tan \theta =\pm \sqrt{1-r^2}$. The DOS is similar to a state with linear line nodes. For $r=1$, there is a pair of quadratic nodes along the $\hat{x}$-axis with linear DOS. The DOS for other FM nonunitary states remains qualitatively same.

Now, we look at the residual thermal conductivity for the two component states on a spherical Fermi surface. Fig. \ref{fig:k0_sph} shows the normalized thermal conductivity in the zero temperature limit for the $A_{1u}+irB_{1u}$ state in panels (a) to (c). For the gapped state ($r<1$), this state shows complete isotropic behavior and the $\hat{z}$-axis remains very closer  to the $xy$-plane thermal response. For $r=1$, this has nodes in the $xy$-plane, therefore, as a function of disorder, $\hat{z}$-axis is weaker, but as the disorder level increases, the $\hat{z}$-axis response dominate, because there are minima along the $xz$ and $yz$ planes, and $xy$-plane also contains a maxima. For $r>1$, there are quadruple pairs of point nodes in the $xz$ and $yz$planes, therefore, $\hat{z}$-axis always dominates. For  the $A_{1u}+irB_{2u}$  and the $A_{1u}+irB_{3u}$ states,  $\hat{z}\leftrightarrow \hat{y}$ and $\hat{z}\leftrightarrow \hat{x}$, respectively, in  Fig. \ref{fig:k0_sph}(a) to (c).

Next, we consider the $B_{2u}+irB_{3u}$ state, which is a chiral state on a spherical Fermi surface.  Fig. \ref{fig:k0_sph} (d) to (f) show the zero temperature limit normalized thermal conductivity. For $r<1$, the nodes are located in the $yz$-plane, therefore, the $\hat{x}$ direction remains smallest. For small values of $r$, nodes are closer to the $\hat{y}$-axis, hence this direction show the largest thermal conductivity. As $r$ increases and reaches closer to unity, the thermal conductivity along the  $\hat{z}$-axis starts to dominate. For $r=1$, there are quadratic point nodes along the $\hat{z}$-axis, the thermal conductivity is largest along the nodal directions. For $r>1$, nodes are located in the $xz$-plane, and the thermal conductivity becomes smallest along $\hat{y}$ direction. For values closer to unity, $\hat{z}$-direction shows highest normalized thermal conductivity, but as the $r$ increases, the nodes move towards $\hat{x}$-axis, which shows highest normalized thermal conductivity.

Finally, we look at the temperature dependence of the normalized thermal conductivity for the $A_{1u}+irB_{1u}$, which is shown in Fig. \ref{fig:kt_sph}(a) to (f). For the $A_{1u}+irB_{1u}$, the anisotropy as a function of temperature follows the zero temperature trend. in Fig. \ref{fig:kt_sph}(g) to (l) show the normalized thermal conductivity for the $B_{2u}+irB_{3u}$ state, the zero temperature anisotropy continues at low temperature, but at higher temperature the anisotropy reverses as seen in the case of cylindrical Fermi surface due to the upper quasiparticle energy band.
\clearpage
\bibliographystyle{apsrev4-1} 

\end{document}